\documentclass[draft]{agujournal2019}
\usepackage{url} 
\usepackage{lineno}
\usepackage[inline]{trackchanges} 
\usepackage{soul}

\draftfalse

\journalname{Earth and Space Science}

\begin{document}

%
%


\title{The boulder population of asteroid 4 Vesta: Size-frequency distribution and survival time}

%
%




\authors{Stefan E. Schr\"oder\affil{1}, Uri~Carsenty\affil{1}, Ernst~Hauber\affil{1}, Franziska~Schulzeck\affil{1}, Carol~A.~Raymond\affil{2}, Christopher~T.~Russell\affil{3}}

\affiliation{1}{Deutsches Zentrum f\"ur Luft- und Raumfahrt (DLR), 12489 Berlin, Germany}
\affiliation{2}{Jet Propulsion Laboratory (JPL), California Institute of Technology, Pasadena, CA 91109, U.S.A.}
\affiliation{3}{Institute of Geophysics and Planetary Physics (IGPP), University of California, Los Angeles, CA 90095-1567, U.S.A.}




\correspondingauthor{Stefan Schr\"oder}{stefanus.schroeder@dlr.de}




\begin{keypoints}
\item We mapped boulders larger than 60~m on asteroid Vesta and found all associated with impact craters
\item The maximum lifetime of these large Vesta boulders is about 300~Ma, similar to that of meter-sized lunar boulders
\item Their cumulative size-frequency distribution is best fit by a Weibull distribution rather than a power law
\end{keypoints}

%
%

%
%


\begin{abstract}

Dawn's framing camera observed boulders on the surface of Vesta when the spacecraft was in its lowest orbit ({\sc lamo}). We identified, measured, and mapped boulders in {\sc lamo} images, which have a scale of 20~m per pixel. We estimate that our sample is virtually complete down to a boulder size of 4~pixels (80~m). The largest boulder is a 400~m-sized block on the Marcia crater floor. Relatively few boulders reside in a large area of relatively low albedo, surmised to be the carbon-rich ejecta of the Veneneia basin, either because boulders form less easily here or live shorter. By comparing the density of boulders around craters with a known age, we find that the maximum boulder lifetime is about 300~Ma. The boulder size-frequency distribution (SFD) is generally assumed to follow a power law. We fit power laws to the Vesta SFD by means of the maximum likelihood method, but they do not fit well. Our analysis of power law exponents for boulders on other small Solar System bodies suggests that the derived exponent is primarily a function of boulder size range. The Weibull distribution mimics this behavior and fits the Vesta boulder SFD well. The Weibull distribution is often encountered in rock grinding experiments, and may result from the fractal nature of cracks propagating in the rock interior. We propose that, in general, the SFD of particles (including boulders) on the surface of small bodies follows a Weibull distribution rather than a power law.

\end{abstract}


%
%

\section{Introduction}

Boulders on small Solar System bodies provide a window into the interior. Boulders may be created by spallation during large impacts and therefore are typically found in and around fresh craters. They do not survive forever, but are gradually eroded into dust by exposure to the space environment \cite{D14,B15}. In the literature, the location and outline of boulders are typically mapped and a size-frequency distribution (SFD) is produced. Studies of main-belt asteroids often focus on finding the crater of origin of sparsely distributed boulders \cite{L96,T01,K12}. In contrast, the surface of near-Earth asteroids is densely populated by boulders of all sizes, all thought to originate from the destruction of a parent body \cite{M14,DG19,M19}. Generally, a power law is fitted to the cumulative boulder SFD, with an exponent defining the slope. In a seminal paper, \citeA{H69} provided a table in which empirically-derived exponents are linked to various geological processes. The largest exponent in the table is associated with the ejecta of hypervelocity impacts. The exponents of asteroid boulder SFDs are generally found to be close to this value. However, published exponents show considerable variation, and the reason for this is often not clear. What clues does a particular value of the exponent provide to the composition and physical properties of the surface?

From its vantage point in the lowest mapping orbit, the {\sc nasa} Dawn spacecraft was able to distinguish boulders on the surface of Vesta. A first analysis of the global boulder population was performed by \citeA{D16}, who inferred a typical regolith depth of about 1~km. We extend their analysis by mapping the location of all boulders that we could recognize to investigate their distribution in and around individual craters, as well as the distribution of craters with boulders over the surface. We estimate the average boulder lifetime by comparing the boulder density around craters for which an age estimate is available \cite{SK14,K14,R14}, and assess the \citeA{B15} prediction that meter-sized boulders on Vesta live roughly 30~times shorter than on the Moon. We also determined the SFD of boulder populations of individual craters and that of the global population. Vesta is a very large asteroid, and on its surface we found boulders with a size of hundreds of meters, larger than those found on other asteroids, with the exception of Ceres \cite{S18}. Such huge objects are also rare on Earth \cite{BR17}, and we will therefore explore a relatively unknown part of the boulder SFD. The official geological term for terrestrial particles larger than a few meters is ``megaclast''. According to \citeA{BR17}, our Vesta megaclasts classify as megablocks ($10 < d < 100$~m) and superblocks ($d > 100$~m). In this paper we will continue to refer to all Vesta megaclasts as ``boulders'', for convenience. We compare the Vesta boulder SFD with that of other small bodies. This is made challenging by the fact that different fitting methods have been used in the literature to fit the power law, only one of which, the maximum likelihood (ML) method, is statistically correct \cite{C09}. We therefore carefully assess published exponents, and in some cases reanalyze the original data using the ML method.

In this paper we only discuss boulders on small bodies. We do not discuss boulders on Earth and Mars, which are often formed and weathered by forces other than those imposed by the space environment (tectonic, aeolian, biological) \cite{P17}. Neither do we consider boulders on comets, as they are not thought to result from impacts but rather from processes like sublimation, activity outbursts, and the associated uplifting and re-deposition \cite{P15}.

\section{Methodology}

\subsection{Boulder mapping}

Boulders on Vesta can only be distinguished in framing camera images from the Low Altitude Mapping Orbit ({\sc lamo}), which were acquired between 13 December 2011 and 30 April 2012 at an altitude of 180~km \cite{R07}. The framing camera is a narrow-angle camera with a field-of-view of $5.5^\circ \times 5.5^\circ$ \cite{S11}. {\sc lamo} images have the highest resolution of all images acquired at Vesta. Most {\sc lamo} images have a spatial resolution between 17 and 22~m, although south of latitude $-45^\circ$ the resolution is a little lower \cite{R13}. When we talk about ``pixels'' in this paper, we always refer to {\sc lamo} pixels. We selected 53~{\sc lamo} images that we used in our analysis (at least one for each crater with boulders), and determined their average spatial resolution as $20 \pm 2$~meter per pixel. In this paper, we therefore adopt a typical {\sc lamo} image resolution of 20~m per pixel. We are confident that we could reliably identify a boulder for a size of at least 3~pixels (60~m) using the method described below, although we mapped suspected boulders smaller than that. However, we cannot assume our count is complete with the 3~pixels criterion, if only because of the measurement uncertainty. A criterion of 4~pixels (80~m) is more likely to ensure near-completeness \cite{P16}. The photometric angles at the center of the 53 images are plotted as a function of latitude in Fig.~\ref{fig:viewing_angles}. The illumination and observation conditions during {\sc lamo} were relatively constant. The average photometric angles at the center of the images are: incidence $\iota = 57^\circ \pm 10^\circ$, emission $\epsilon = 9^\circ \pm 4^\circ$, and phase $\alpha = 54^\circ \pm 8^\circ$. The incidence and phase angles become relatively large only towards the mid-latitudes in the northern hemisphere ($+50^\circ$), which were partly in the shadow. This means that the majority of boulders were observed under similar conditions.

The second author reviewed the entire {\sc lamo} data set and identified, measured, and mapped all boulders using the J-Vesta GIS program, which is a version of {\sc jmars} ({\url{https://jmars.mars.asu.edu/}) \cite{CE09}. The first author reviewed these results for accuracy and completeness. The location of boulders (and craters) is defined in the Claudia coordinate system \cite{Ru12}. Boulders were identified as positive relief features in projected images at a zoom level of 1024~pixels per degree. The native {\sc lamo} resolution is $\sim 200$~pixels per degree, so this represents a zoom factor of about 5. The boulder size was determined using the J-Vesta crater measuring tool, which draws a circle around the boulder fitted to 3~points that are selected by the user on the visible boulder outline. The vast majority of boulders that we mapped were too small to clearly distinguish the shape, and we therefore chose the circle as a reasonable approximation. The first author also mapped boulders for a single crater using the same method to confirm that the measurement uncertainty is about a single pixel. While mapping, there were several challenges to overcome. The limited accuracy of pointing information for {\sc lamo} images leads to mismatches between projected images. We used small craters inside and outside the crater as tie points to align the projected images to the Vesta background mosaic, and to align images relative to each other. J-Vesta allows shifting the projected image relative to the background in the horizontal and the vertical direction, but rotation is not possible. All this leads to uncertainty in the location of boulders on the order of 500~m. In addition, unusually bright inner walls of fresh craters could appear saturated, making it impossible to recognize any feature on their surface, due to a limitation in J-Vesta with image brightness scaling. Generally, on crater walls it was difficult to distinguish genuine boulders from rocky outcrops, which are prevalent just below the rim. When in doubt, we did not include such features in our sample. It was also difficult to decide whether a large mound was a degraded boulder or had always been simply a pile of rubble. The north pole of Vesta was in the shadow at the time of {\sc lamo} (December 2011 to April 2012). Therefore counts of boulders around craters are incomplete north of latitude $+30^\circ$, and it was not possible to identify boulders north of $+60^\circ$.

\subsection{Size-frequency distribution}

Boulder SFDs are often displayed in cumulative format, following the recommendation by the \citeA{C79} for crater SFDs to be plotted in both the cumulative and differential format. For the latter, the Working Group recommended a relative distribution known as the R-plot. In this paper, we display the boulder SFD in both cumulative and differential format, but choose the incremental (binned, histogram) version for the latter \cite{C93}. Figure~\ref{fig:generic_acuracy} illustrates both formats for a simulated boulder population.

The cumulative format has become especially popular in the literature (perhaps because of its remarkable ability to convert even the noisiest data into a smooth downward curve), and figures in differential format are often omitted from boulder (and crater) counting papers. Several unfortunate practices associated with the cumulative distribution have become established in the literature. The first is binning. The \citeA{C79} wrote that the ``collection, manipulation, and display of unbinned data is more time consuming than for binned data''. This may have been true in 1979, but is no longer the case. Binning of a cumulative distribution is not merely unnecessary, but represents a loss of information. Also, bins that have the same value as their neighbor on the right are often not displayed. Their omission skews the appearance of the distribution and affects how we perceive the goodness-of-fit of a model curve. Another statistically suspect practice is the association of Poisson (square-root) error bars with the bins. Such error bars are valid for single bins when considered in isolation, but in a plot of the cumulative distribution they do not serve their usual purpose of indicating the uncertainty of the data: Regardless of the size of its bar, the number in a bin can neither be lower than its neighbor on the right nor higher than its neighbor on the left. Therefore, we do not bin the cumulative distribution in this paper. The differential distribution is always binned in practice, where the bin width can be chosen as constant on a linear or logarithmic scale. Poisson error bars associated with the bins are statistically meaningful and will give the correct impression of how well the data agree with a particular model curve. Empty bins, however, present a problem. SFD plots are invariably shown on a log-log scale, and empty bins cannot be displayed, which skews the appearance of the distribution. Moreover, empty bins cannot be included when fitting models (such as a power law) to the logarithm of the data, introducing bias.

\subsection{Power law fitting}

The cumulative distribution of boulders on Solar System bodies is generally assumed to follow a power law. The number of boulders with a size larger than $d$ is:
\begin{equation}\label{eq:ML}
N(>d) = N_{\rm tot} \left( \frac{d}{d_{\rm min}} \right)^\alpha,
\end{equation}
with $\alpha < 0$ the power law exponent and $N_{\rm tot}$ the total number of boulders larger than $d_{\rm min}$. \citeA{H69} showed that the exponent of a cumulative distribution of a quantity that follows a power law is identical to that of the associated incremental differential distribution with a constant bin size on a logarithmic scale. \citeA{C93} added that this is only true if the logarithmic bins are chosen wide enough. If the bin size of the incremental differential distribution is constant on a linear scale, then the exponent differs by unity from that of the cumulative distribution. Historically, the power law exponent was estimated by fitting a line to the cumulative distribution plotted on a log-log scale, either by eye or by means of a least-squares algorithm. There are two problems associated with this practice. First, the differential distribution may clearly show that a certain quantity does not follow a power law, but the associated cumulative distribution may still give the impression that it does. Second, the uncertainty of the power law exponent cannot be reliably retrieved by means of simple linear regression because of the ill-defined errors associated with the cumulative bins. Uncertainties given for exponents derived from a conventional fit to the cumulative distribution are underestimated \cite{C09}. A statistically sound way to estimate the power law exponent from the SFD is the maximum likelihood (ML) method \cite{N05,C09}. In the ML method, the power law exponent is calculated directly from the boulder size measurements, and is therefore independent of how the data are displayed (cumulative, differential, binning). The ML estimate for the power law exponent ($\alpha < 0$) is:
\begin{equation}\label{eq:ML_exp}
\hat{\alpha} = -N \left( \sum_{i=1}^{N} \ln \frac{d_i}{d_{\rm min}} \right)^{-1},
\end{equation}
with $d_i$ the size of boulder $i$ and $N$ the total number of boulders with a size larger than $d_{\rm min}$. The standard error of $\hat{\alpha}$ is
\begin{equation}\label{eq:ML_err}
\sigma = -\hat{\alpha} / \sqrt{N}
\end{equation}
plus higher order terms, which we ignore in this paper. We note that the cumulative power law exponent alpha ($\alpha_{\rm S} < 0$) in Eq.~\ref{eq:ML_exp} relates to the scaling parameter alpha ($\alpha_{\rm C} > 0$) in \citeA{C09} as $\alpha_{\rm S} = 1 - \alpha_{\rm C}$. The estimator in Eq.~\ref{eq:ML_exp} is unbiased only for sufficiently large sample size; \citeA{C09} suggest $N > 50$. \citeA{C09} also provide details to a statistical test that evaluates whether a power law is an appropriate model for the data. The test randomly generates a large number of synthetic data sets according to the best-fit power law model (specified by $\hat{\alpha}$ and $d_{\rm min}$), and calculates for each the so-called Kolmogorov-Smirnov statistic, which is a measure of how well the synthetic data agree with the model. A $p$-value, defined as the fraction of synthetic data sets that have a larger statistic than the real data set, quantifies how well the power law performs. The authors adopt $p < 0.1$ to mean rejection of the power law model.

An example of a fit with the ML method is shown in Fig.~\ref{fig:generic_acuracy}A. In this paper, we generally display the power law associated with a ML-derived exponent in plots of both the cumulative and differential distributions. Displaying it in the latter format requires an additional step of performing a conventional fit to the differential distribution to estimate the intercept in addition to the exponent. In case the individual boulder sizes are not available, the ML method cannot be applied. If binned boulder numbers as a function of size are available, fitting a power law to the incremental differential distribution in a log-log plot by means of a least-squares algorithm is preferred over fitting a power law to the cumulative distribution. Still, one must choose an appropriate bin size and the necessary exclusion of empty bins (the logarithm of zero is undefined) introduces bias, as we will quantify below. One could resolve the problem of empty bins by fitting the power law to the data on a linear scale, but this gives unduly weight to bins with larger numbers towards smaller boulder sizes, which runs counter to the purpose of the power law (describing identical behavior over a wide range of scales). An example of a power law fit to the differential distribution is shown in Fig.~\ref{fig:generic_acuracy}B.

\subsection{Monte Carlo simulations}

To investigate the statistical power of the different methods to retrieve the power law exponent (ML and differential fit) and to uncover any associated bias, we simulate the process by randomly generating power law distributions. Fitting a power law to the cumulative distribution is such poor practice that we do not evaluate the method here. The continuous power law probability distribution is also known as the Pareto distribution \cite{N05}. To simulate a size distribution of boulders associated with impact craters we draw a random variate $U$ from a uniform distribution on $(0, 1)$ using the {\sc randomu} routine in IDL with an undefined seed. Then the boulder diameter
\begin{equation}\label{eq:Pareto}
d = d_{\rm min} U^{1/\alpha}
\end{equation}
follows a Pareto distribution, with $\alpha$ the power law index (associated with the cumulative distribution function, with $\alpha < 0$) and $d_{\rm min}$ the minimum boulder diameter. We start our investigation by generating 150 boulder populations with the number of boulders in each chosen randomly in the logarithmic interval $(10, 1000)$. For each population, the number of boulders served as input to Eq.~\ref{eq:Pareto} to generate a SFD that obeys a power law with a (cumulative) exponent of $-4.0$. Then we estimate the power law exponent of each simulated population by means of two alternative methods: the ML method and a conventional (least-squares) power law fit to the differential distribution in log-log format. The ML method estimates the power law exponent straight from the boulder counts, but when fitting the differential distribution one must make several choices. For practical purposes we adopt the parameters used in this paper to represent Vesta: a minimum boulder size of $d_{\rm min} = 80$~m and a constant bin width of 0.07 on a logarithmic scale. This corresponds to $\beta = 10^{0.07} - 1 = 0.17$ and meets the bin width criterion of \citeA{C93}, so that the derived exponent should be the same as that of the cumulative distribution. We exclude any filled bins beyond the first empty one from the fit, which introduces bias in the estimated exponents, especially for small boulder populations (simply excluding empty bins from the fit introduces a different bias). Fitting was performed using the Levenberg-Marquardt algorithm with constrained search spaces for the model parameters \cite{M78,Ma09}. The Poisson errors on the logarithm of the binned counts are asymmetric, and we pass the logarithm of the upper error to the fitting routine.

The results of the simulation are shown in Figs.~\ref{fig:generic_acuracy}C and D. For both methods, a large number of boulders is required to reliably estimate the power law exponent. But even for 1000 boulders the estimated exponent may differ from the true value ($-4.0$) by up to 0.3, simply by chance. For 100 boulders the estimate exponent is typically off by unity. Below 100 boulders the estimated exponent may rapidly diverge from the true value. For more than 200~boulders, the two methods give similar results. For smaller numbers, however, the estimated exponents are not distributed symmetrically around the true value. For the ML method this is the bias indicated by \citeA{C09}. It appears to be negative, but the situation is more complicated than Fig.~\ref{fig:generic_acuracy}C suggests. The median exponent is actually $-4.0$, meaning that exponents smaller and larger than $-4$ are found in roughly equal numbers. However, compared to the true value, the estimated exponent can be much more negative (down to $-9$) than less negative (up to $-2$). This is reflected in the mean of the exponents in the figure, which is $-4.2$. For the method of fitting the differential distribution (Fig.~\ref{fig:generic_acuracy}D) the median and mean of the exponents are $-3.6$ and $-3.5$, respectively. This positive bias results from the aforementioned exclusion of empty bins, and affects boulder populations of all sizes. Therefore, the differential fit method is less accurate than the ML method.

We find that the derived power law exponent for populations with a small number of boulders is always biased, but more so for the differential fit method than for the ML method. For the latter, we recommend a population size of at least 100~boulders for which the size is accurately known (at least 4~image pixels). Here we are more conservative than \citeA{C09}, who recommended a minimum sample size of 50. For larger boulder numbers, the estimated exponent generally differs from the true value by less than unity. But even with 1000 boulders, we can still expect differences of up to 0.3. This is important to keep in mind, for example, when we consult \citeA{H69} for the interpretation of the power law exponent. The author provided a list of exponents associated with fragmented rocks created by different geological processes with an accuracy of two decimal numbers. We verified that an accuracy of 0.01 for the exponent is only derived from populations of at least a million fragments of known size.

\subsection{Weibull distribution}

Although the power law enjoys widespread use as a model for the boulder SFD, it does not always fit the data well. Several authors used exponential functions to fit the SFD of boulders on Mars \cite{GR97,G03,P17}, and \citeA{P19} found the Weibull distribution to fit the SFD of boulders around a lunar crater better than a power law. Exponential functions can be considered as variations of the Weibull distribution, so it is worth considering the latter as a viable alternative to the power law.

The Weibull distribution was initially derived empirically, and is often used to describe the particle distribution resulting from grinding experiments \cite{RR33}. Seeking to describe such a distribution, \citeA{B89} developed a theory of sequential fragmentation. He defined the particle distribution as
\begin{equation}\label{eq:seq_frag}
n(m) = C \int_{m}^{\infty} n(m^\prime) f(m^\prime \rightarrow m) \mathop{d m^\prime},
\end{equation}
where $n(m)$ is the number of particles per unit mass of mass between $m$ and $m + dm$. The function $f$ describes the mass distribution that results when a single fragment of mass $m^\prime > m$ breaks into smaller, lighter pieces, and takes the form of a power law:
\begin{equation}\label{eq:single_particle_frag}
f(m^\prime \rightarrow m) = \left( \frac{m}{m_1} \right)^\gamma,
\end{equation}
with exponent $-1 < \gamma < 0$ and $m_1$ a scaling factor related to the average mass in the distribution $n(m)$. The mass on the right hand side of Eq.~\ref{eq:single_particle_frag} is $m$ and not $m^\prime$. \citeA{BW95} showed that a power law follows naturally from a single-event fragmentation that leads to a branching tree of cracks that have a fractal character. The spacing of the cracks is described by the fractal dimension $D_{\rm f} = -3 \gamma$. The solution of Eq.~\ref{eq:seq_frag} is a Weibull distribution:
\begin{equation}\label{eq:Weibull_mass_diff}
n(m) = \frac{N_{\rm T}}{m_1} \left( \frac{m}{m_1} \right)^\gamma \exp \left[ - \frac{(m / m_1)^{\gamma + 1}}{\gamma + 1} \right],
\end{equation}
with Weibull shape parameter $\gamma + 1$. The cumulative form of Eq.~\ref{eq:Weibull_mass_diff} is given by:
\begin{equation}\label{eq:Weibull_mass}
N(>m) = N_{\rm T} \exp \left[ - \frac{(m / m_1)^{\gamma + 1}}{\gamma + 1} \right],
\end{equation}
For use in this paper, we convert Eq.~\ref{eq:Weibull_mass} to an expression for the cumulative number density as a function of particle diameter $d$:
\begin{equation}\label{eq:Weibull_size}
N(>d) = N_{\rm T} \exp \left[ - \frac{(d / d_1)^{3(\gamma + 1)}}{\gamma + 1} \right],
\end{equation}
using $m / m_1 = (d / d_1)^3$, with $d_1$ a size scaling constant. The Weibull shape parameter in Eq.~\ref{eq:Weibull_size} is $3(\gamma + 1)$.
In practice, we only include boulders larger than a certain size in the analysis. That means that we are dealing with a left-truncated Weibull distribution with the cumulative form \cite{W89}:
\begin{equation}\label{eq:Weibull_min_size}
N(>d) = N \exp [ - \alpha ( d_i^\beta - d_{\rm min}^\beta ) ],
\end{equation}
where $N$ is the number of boulders larger than $d_{\rm min}$. Often used to define the Weibull distribution are the scale parameter $\lambda = \alpha^{-1/\beta}$ and shape parameter $k = \beta = 3(\gamma + 1)$. We estimate $\alpha$ and $\beta$ from the boulder sizes $d_i > d_{\rm min}$ using the ML method. To maximize the log-likelihood function, these two equations must be satisfied:
\begin{eqnarray} \label{eq:Weibull_estimation}
\alpha = \frac{N}{\sum (d_i^\beta - d_{\rm min}^\beta)} \nonumber \\
\frac{N}{\beta} + \sum \ln d_i - N \frac{\sum (d_i^\beta \ln d_i - d_{\rm min}^\beta \ln d_{\rm min})}{\sum (d_i^\beta - d_{\rm min}^\beta)}= 0.
\end{eqnarray}
We find $\hat{\beta}$ from a simple grid search, and $\hat{\alpha}$ by inserting $\hat{\beta}$. The Weibull distribution is also discussed by \citeA{C09} as the ``stretched exponential'' distribution.

\section{Results}

\subsection{General statistics}

Boulders on Vesta come in many shapes and sizes. In total, we identified 6577 boulders on the surface of Vesta with a diameter larger than 3~image pixels (60~m), of which 2318 were larger than 4~pixels (80~m). We found that all boulders are associated with impact craters, whose diameter we indicate with $D$. The details of all craters with at least one boulder larger than 4~pixels ($n = 69$) are listed in Table~\ref{tab:craters}. Examples of Vesta boulder morphology are shown in Fig.~\ref{fig:boulders}, which also illustrates one of the challenging aspects of boulder identification: Boulders change shape over time. Young craters have boulders that are easily recognized with their well-defined shadows. Old craters have only few boulders, most of which are large and rather look like mounds, with diffuse shadows. For such craters it is likely that the more numerous small boulders have simply degraded beyond recognition. We will discuss the craters in Fig.~\ref{fig:boulders} in more detail in Sec.~\ref{sec:age}.

We summarize the general statistics of the global boulder population in Fig.~\ref{fig:basic_stats}. The number of boulders per crater clearly depends on crater size (Fig.~\ref{fig:basic_stats}A). The correlation is not very tight because the number also depends on crater age. The size of the largest boulder of a crater (diameter $L$) depends on the crater size too (Fig.~\ref{fig:basic_stats}B). In contrast to the number of boulders, the size of the largest boulder does not necessarily depend on age, as boulders may disintegrate in place to form equally-sized mounds (Fig.~\ref{fig:boulders}). Being a single measurement, the largest boulder size is a poor statistic, but is nevertheless often used to characterize boulder populations. The largest boulder we identified on Vesta is around 400~m large and located on the floor of Marcia crater. The inset in Fig.~\ref{fig:basic_stats}B shows the challenges of assigning a unique size to large boulders, which often have an irregular shape. In the figure we simply assumed a 1~pixel measurement uncertainty, but the actual uncertainty increases with boulder size. Boulders of such huge size are unknown on Earth, but their identification on Vesta is unambiguous. \citeA{L86} provided a plot of the size of the largest boulder on the Moon and the Martian moons Deimos and Phobos. The Vesta largest boulder distribution agrees well with theirs where our crater size ranges overlap ($3 < D < 10$~km). In Fig.~\ref{fig:basic_stats}B we also compare our distribution with the relation provided by \citeA{L96} ($L = 0.25 D^{0.7}$ with $L$ and $D$ in m) and with the empirical range established by \citeA{M71} for a selection of lunar and terrestrial craters ($L = 0.01^{1/3} K D^{2/3}$ with $K$ ranging from 0.5 to 1.5). The former relation represents more or less the upper limit of the latter range. We find that the largest boulders of the smaller craters ($D < 10$~km) agree well with the \citeA{L96} relation, whereas the largest boulders of the larger craters ($D > 10$~km) agree better with the \citeA{M71} range. But our total population of largest boulders does not agree with either relation. How uncertain are our size measurements? The first author of this paper carefully verified all largest boulders as identified by the second author. For example, there is indeed a 220~m sized large, angular, block on the rim of the 7.5~km sized crater Unnamed12. Of course, we cannot conclude with certainty that this superblock is indeed a (former) boulder ejected by the impact that created the crater. We also note that the second largest block around Unnamed12 is less than half the size of the largest, which is more in line with the aforementioned relations. Given the uncertainties in our method and the empirical relations, we conclude that the largest boulder sizes on Vesta agree reasonably well with what has been observed on other small bodies.

\subsection{Spatial distribution}

In Fig.~\ref{fig:global}, we plot the distribution of all boulders on the surface of Vesta with a size of at least 3 pixels on an albedo map. Our map differs from the craters-with-boulders map of \citeA{D16}, because of our restriction on boulder size and their restriction on crater size ($<12$~km diameter). The largest crater on Vesta with boulders, Marcia, dominates the center of the map. Even though the count of 3~pixel sized boulders is most likely incomplete (see Sec.~\ref{sec:Vesta_size-freq-dist}), we nevertheless chose to display these instead of 4~pixel sized boulders, as there are relatively few of the latter. First we note that the boulder count is incomplete for craters at high latitudes in the northern hemisphere, because their floors were mostly in the shadow during {\sc lamo}. South of $50^\circ$N, the spatial distribution of craters with boulders appears not to be entirely random. The albedo map gives the impression that boulders tend to avoid a large area of below-average albedo, color-coded blue, which was also noticed by \citeA{D16}. The dark material in this area may have been delivered by a large carbonaceous chondrite impactor that created the ancient Veneneia basin \cite{R12,P12,J14}. The ejecta of the other large basin on the south pole, Rheasilvia, are distributed over the entire southern hemisphere \cite{Y14}. Craters with boulders dot the southern hemisphere and there is no obvious north-south gradient in the abundance of such craters. Therefore, the scarcity of craters with boulders on dark terrain may be related to the, presumably carbonaceous, composition, or the fact that these are former ejecta. Boulders may form less easily here, \citeA{D16} suggested that the regolith is thicker than average, or live shorter. These ideas can be tested by studying the boulder population of Ceres, Dawn's next mission target, whose reflectance spectrum is similar to that of carbonaceous chondrites \cite{McCG74}. Marcia, which is located within the dark terrain, has abundant boulders. As the dark layer is relatively thin \cite{J14}, the impactor that formed Marcia likely punched through it, and the boulders are composed of the underlying, non-dark, material.

The distribution of boulders in and around the craters on Vesta is highly variable. We show the boulders distribution for a selection of craters in Fig.~\ref{fig:projections}. Some craters have most boulders located outside the rim (Cornelia, Licinia), others have most boulders located inside the crater (Marcia, Vibidia). Boulders associated with craters that formed on a slope \cite{Kr14} are concentrated on the down-slope side of the crater (Antonia, Unnamed36). We plot the boulder locations in two colors: green for boulders with a size between 3 and 4~pixels, and red for boulders larger than 4 pixels. The figure shows that boulders of this size ($d > 60$~m) are generally found within one crater diameter of the rim. Rarely did we find boulders further away; in such cases the identification was invariably ambiguous. We expect that large boulders are found closer to the crater rim than smaller boulders because of the larger amount of energy required to eject them, a prediction confirmed for lunar craters \cite{BM10,K16,P19}. However, the red and green boulders in Fig.~\ref{fig:projections} are not clearly segregated in terms of distance from the crater center. A clear correlation between boulder size and distance from the crater was not found for Ceres either \cite{S18}. But it is still possible that such a correlation exists for Vesta, and that we failed to find it only because boulders smaller than 60~m could not be reliably identified. The number of boulders around each crater not only depends on the crater size, but also on age. We discuss the latter dependency in the following section.

Figure~\ref{fig:projections} does not distinguish between boulders in- and outside the crater rim. We found little evidence for the production of boulders by physical weathering from crater walls. Rock falls can be triggered by thermally-induced mechanical stresses due to varying insolation \cite{AV13}. However, we could not identify any boulder tracks on the talus material that mantles many inner crater walls, in contrast to the Moon where boulder tracks are common \cite{SK16,B19}. We also analyzed the distribution of boulders with respect to the local slope. The boulder density is low on steep slopes, as expected, but there is no systematic increase of boulder density at the foot of steep slopes, which would be expected if physical weathering recently produced new boulders here. This lack of evidence for the production of boulders larger than 60~m by post-impact weathering is consistent with the notion of \citeA{D16} that boulders on Vesta are excavated from bedrock or regolith by impacts. As boulders on Vesta appear to be the direct result of a single geologic process, i.e.\ impact cratering, we do not distinguish boulders located in- and outside of craters in the remainder of this paper. This also resolves the complication that ``inside'' and ``outside'' are poorly defined for craters that formed on a slope \cite{Kr14}.

\subsection{Boulder lifetime}
\label{sec:age}

Boulders degrade over time and eventually disappear from the surface. Two mechanisms thought to be responsible for boulder decay on asteroids are collisional fragmentation by meteorite impact and thermal fatigue \cite{D14,B15}. On Vesta we expect collisional fragmentation to dominate because the consequences of the diurnal cycle are probably too small, mainly because of the large distance from the Sun \cite{MB12,B15,MB17}. By comparing the boulder abundance with a crater's age, we can estimate the typical boulder survival time. \citeA{B15} predicted that the boulder survival time on Vesta is much smaller than on the Moon, based on estimates of the potential impactor flux and the expected impact velocities. The expected boulder survival time scales with the inverse of the impact energy, which scales as the impactor flux times the square of the impact velocity. The authors used all known orbits of minor bodies in the inner Solar System to estimate the flux of potential boulder-destroying impactors, which are too small to be observed. The impactor flux predicted in this way for Vesta is about 300 times larger than that for the Moon. The typical impact velocity on Vesta is about a third of that on the Moon. So, the estimated boulder survival time on Vesta is about $9 / 300 = 0.03$ times that on the Moon, i.e.\ 30 times shorter. Age estimates derived from crater counting are available for some Vesta craters with boulders, so we can test this hypothesis.

Two alternative chronologies exist for Vesta \cite{W14b}: the model of \citeA{SK14}, based on the lunar-derived crater production and chronology functions, and the \citeA{MB14,OB14} model, with crater production and chronology functions derived from models of asteroid belt dynamics. For this paper we accept the \citeA{SK14} chronology, not because we necessarily believe it is to be preferred over the other, but because more ages based on this model are available in the literature. Table~\ref{tab:ages} lists the estimated ages of the craters shown in Fig.~\ref{fig:projections}, together with the areal density of their boulders. The ages span a range of 2~Ma to more than 300~Ma. For these craters, the \citeA{SK14} chronology yields lower ages than the \citeA{MB14,OB14} chronology \cite{K14}, so the tabulated ages are conservative estimates. The tabulated areal density is defined as the total number of boulders identified in and around a crater divided by the crater area, which is calculated as the area of a circle with the diameter for that crater. Figure~\ref{fig:density_vs_age} shows how the boulder density varies with crater age. Assuming that the initial boulder density is similar for all craters, the figure confirms the expected correlation of density with age. The figure is for boulders with a minimum size of 3~pixels, but the results are similar for a minimum size of 4~pixels. Some data points in the figure are less reliable than others. The boulder count is incomplete for craters at high latitudes in the northern hemisphere, and craters whose boulder densities are consequently underestimated are Arruntia, Mamilia, and Scantia (open symbols in Fig.~\ref{fig:density_vs_age}A). Crater Unnamed36 has a much higher boulder density than expected for its age of about 250~Ma. This crater is an outlier for two possible reasons: First, the age estimate is based on a small number of craters ($<50$) and may not be fully reliable for the same reason that power law exponents derived from boulder populations of small size are not reliable. Second, it is not clear whether some of the larger mounds that we identified as degraded boulders were once truly boulders or have always been mounds of rubble, as the high abundance of suspected boulders may be related to the fact that this crater formed on a slope \cite{Kr14}. The oldest crater for which we identified boulders is Oppia, with an age of $320 \pm 24$~Ma \cite{SK14}. We did not find boulders around Octavia crater ($147^\circ{\rm E}, -3^\circ$), whose age is estimated as $390 \pm 28$~Ma by \citeA{SK14} and 280-360~Ma by \citeA{W14a}. Then, the maximum age of large boulders ($> 60$~m) appears to be around 350~Ma.

The Vesta boulder density-age relation compares well to that for boulders on the Moon shown in \citeA{B15}, who wrote that after ``a few million years, only a small fraction of meter-sized [lunar] boulders are destroyed but after several tens of million years $\sim 50$\% are destroyed, and for times of 200-300 Ma, $\sim 90$ to 99\% of the original boulder population is obliterated''. Judging from Fig.~\ref{fig:density_vs_age}, this statement also perfectly applies to Vesta boulders, where we note that the scatter in the Vesta data is large, but of the same order as that in the lunar data shown in \citeA{B15}. Because the ages based on the \citeA{SK14} chronology are conservative estimates, we conclude that the boulder survival time on Vesta is at least as long as that on the Moon. The \citeA{B15} prediction appears to be incorrect by more than a factor 30. What can explain the discrepancy? For a start we will assume that Vesta boulders are not more resistant to degradation than lunar boulders given their similar composition, and that both available Vesta chronologies do not dramatically overestimate the crater ages. The impact velocities seem to be uncontroversial, as \citeA{OS11} arrive at similar values as \citeA{B15}. The meteorite flux as estimated by \citeA{B15} from asteroid orbits as published by the Minor Planet Center is very similar for Vesta and Ceres, and about 300 times larger than on the Moon. In a similar exercise, \citeA{OS11} find the impact probability for Ceres to be about 25\% higher than for Vesta because of the former's more central location in the main asteroid belt, but do not provide an impact probability for the Moon. Then the most likely reason for the discrepancy is probably the difference in scale. The lunar boulders considered by \citeA{B15} are typically less than 10~m in size, whereas our Vesta boulders are sized between 60 and 400~m. The implication is that the flux of impactors that can destroy 10~m sized boulders on Vesta is about 30 times larger than the flux of impactors that can destroy 100~m sized boulders.

\subsection{Size-frequency distribution}

\subsubsection{Vesta}
\label{sec:Vesta_size-freq-dist}

First, we pool all boulders counted on the surface of Vesta to find the (cumulative) power law exponent of the global boulder population. Figure~\ref{fig:global_ML} shows the SFD, both in cumulative and differential representation. At the top of the differential plot we show the implications of a 1~pixel measurement error on a logarithmic scale. We chose a logarithmic bin size of 0.07 with the boulder size in meters, ensuring that the size is on the order of the measurement error at the larger end of the scale. As the error is larger than the bin size at the smaller end of the scale, we can expect boulders to spill over into adjacent bins merely by chance. We recognize the characteristic downturn (``roll-over'') of the distributions towards smaller diameters, caused by the limited spatial resolution and the measurement error.

We fit two power laws to the data with the ML method, one with the minimum boulder size ($d_{\rm min}$) fixed, and the other with $d_{\rm min}$ estimated by the ML algorithm. When fixing $d_{\rm min}$ to 4~pixels (80~m), we find a power law exponent of $\alpha = -4.7 \pm 0.1$ ($n = 2319$, black line in Fig.~\ref{fig:global_ML}). By extrapolating this power law to smaller diameters we find that the number of boulders with a diameter of around 3~pixels may be severely underestimated; the observed number in the 3~pixel bin is about 3000, but the extrapolated, expected number is more like 7000. The counts for boulders larger than 4~pixels are probably close to complete. We note that the counts at the largest diameters do not match well with the power law, both in the cumulative and differential representation. Alternatively, when we let the ML algorithm choose the minimum boulder size, we find $d_{\rm min} = 87$~m and a slightly steeper power law with $\alpha = -4.8 \pm 0.1$ ($n = 1575$, red line in Fig.~\ref{fig:global_ML}). The minimum boulder size is close to our earlier estimate of 80~m, confirming that 4~pixels is a reasonable lower limit for the size. But also this power law does not match the counts at large diameters. In fact, the statistical test provided by \citeA{C09} indicates that neither power law in Fig.~\ref{fig:global_ML} is a good model for the data ($p = 0$ and $0.02$, respectively).

We also estimated the power law exponent for each crater individually for craters with at least 6~boulders with a diameter larger than $d_{\rm min} = 80$~m. Plotting these exponents as a function of number of boulders in the population in Fig.~\ref{fig:accuracy}, we find a strong negative bias for smaller populations, which we expected from the generic simulation in Fig.~\ref{fig:generic_acuracy}. We also include three simulations in Fig.~\ref{fig:accuracy} that use the observed population sizes and adopt the best-fit power law exponent for the global boulder population ($\alpha = -4.7$, $d_{\rm min}$ fixed). When we compare the simulations with the observations we notice two things: First, the observed exponents of craters with small boulder populations are typically more negative than in the simulations. Second, the power law exponent of the crater with the largest number of boulders, Marcia, is far from $-4.7$, even though the simulated ``Marcia's'' invariably have exponents close to $-4.7$. This suggests that Marcia's actual exponent significantly differs from that of the global boulder population. We can already see this in Fig.~\ref{fig:projections}, where the ratio of large to small boulders in Marcia is comparatively high. The reason for this is not obvious. There is ample evidence for flows in- and outside the crater \cite{W14a}, so perhaps small boulders were preferentially buried. It is also possible that, because of Marcia's large size, its boulders originate in a deeper, mechanically stronger or less fragmented layer in the interior, making them more resistant to degradation. When we exclude Marcia's boulders from the global population, we derive an exponent of $\alpha = -5.1 \pm 0.1$ ($d_{\rm min} = 80$~m, $n = 1843$, black line in Fig.~\ref{fig:global_ML_noMarcia}), which means a steeper power law. The best-fit power law better fits the observed distributions in the 80-180~m range, although the number of large boulders still appears too low. We repeat our simulations with this revised exponent in Fig.~\ref{fig:accuracy_noMarcia}. Now the observed exponents of craters with small numbers of boulders agree better with those in the simulations. This reinforces the notion that the Marcia boulder population is different from that of all other craters. When we let the ML algorithm estimate the minimum boulder size, we find $d_{\rm min} = 91$~m and an even steeper power law with $\alpha = -5.4 \pm 0.2$ ($n = 1023$, red line in Fig.~\ref{fig:global_ML_noMarcia}). Still, the \citeA{C09} test indicates that neither power law is an acceptable model for the data ($p = 0$ and $0.007$, respectively). Then also without the Marcia boulders, there is no single power law that fits the global distribution over the whole size range.

The failure of the power law to describe the SFD leads us to the Weibull distribution. The question is now whether to include Marcia boulders or not. If the boulder sizes are distributed according to a power law, then exclusion is justified, but the situation is not so clear if we assume a Weibull distribution. We therefore show the best-fit left-truncated Weibull distribution (Eq.~\ref{eq:Weibull_min_size}) in Fig.~\ref{fig:Weibull}, both with (A) and without (B) Marcia boulders. Again we restrict the fit to boulders of size $> 4$~pixels. Including all boulders, the best-fit distribution has $N = 2319$, $\alpha = 2.25$, and $\beta = 0.37$ (scale parameter $\lambda = 0.11$ and shape parameter $k = 0.37$). Without Marcia boulders, the best-fit parameters are $N = 1843$, $\alpha = 0.69$~m, and $\beta = 0.56$ (scale parameter $\lambda = 1.9$ and shape parameter $k = 0.56$). The fractal dimension $D_{\rm f} = 3 - \beta$ for the cracks in the rock is 2.6 and 2.4, respectively. Excluding Marcia boulders changes the Weibull parameters, but does not improve the fit to the data. The Weibull distribution fits the Vesta SFD better than the power law. And contrary to the power law, it does not imply that the number of boulders with a size of 3~pixels is massively underestimated, which seems more reasonable. The Weibull distribution is likely also a better model for the SFD of individual craters. Using the \citeA{C09} test with $d_{\rm min} = 80$~m, we found that the power law fails to fit the SFD of all four craters with more than 100~boulders larger than 4~pixels (Antonia, Licinia, Marcia, and Pinaria).

\subsubsection{Other small bodies}

How does the Vesta boulder SFD compare to those of boulder populations on other small Solar System bodies? In the literature, the boulder SFD distribution is generally assumed to follow a power law. Several authors have now used the ML method to estimate the power law exponent:

{\it 25143~Itokawa.} \citeA{DS15} determined the exponent for the global block distribution of S-type asteroid Itokawa to be $\alpha = -3.6 \pm 0.3$, based on a sample of about 800 boulders from \citeA{M14} in the size range of 7-35~m (using the equivalent spherical radius method for sizing the boulders). \citeA{M19} confirmed this value as $-3.5 \pm 0.2$ in an analysis based on the same measurements. The total number of boulders is large enough to conclude that the Itokawa power law exponent is significantly different from that of Vesta. However, Itokawa is different from Vesta and the other bodies considered in this section, in the sense that its boulders are not associated with particular craters, but are thought to all derive from the disruption of a parent body \cite{M08,N08}.

{\it 1~Ceres.} \citeA{S18} determined the exponent for a number of craters on dwarf planet Ceres and found values in the range of $-6.2$ to $-4.4$. The most reliable value was that obtained for Jacheongbi crater: $\alpha = -4.4 \pm 0.7$, with 147 boulders in the 162-400~m size range, which is not significantly different from the Vesta exponent. More negative values were determined for other craters with fewer boulders, which is consistent with the negative bias at small boulder numbers we observed in our simulations. We pooled all boulder counts from \citeA{S18} (6~craters) and determined the power law exponent using two methods to choose $d_{\rm min}$: (1) setting it to 4~pixels and (2) estimating it by means of the ML algorithm. Figure~\ref{fig:Ceres} show the results in the differential (A) and cumulative (B) representations. For (1) we find an exponent of $-4.3 \pm 0.2$ from measurements of 544 boulders in the 140-394~m size range. For (2) we determined an exponent of $-5.6 \pm 0.4$ from measurements of 159 boulders in the 189-394~m size range. In both cases, the \citeA{C09} test indicates that a power law is not a good model for the data. The shape of the Ceres boulder SFD is similar to that of Vesta in Fig.~\ref{fig:global_ML_noMarcia}, and we suspect that the Weibull distribution may also be a reasonable model for Ceres.

{\it 162173~Ryugu.} \citeA{M19} determined the exponent for the global block distribution of C-type asteroid Ryugu to be $\alpha = -2.65 \pm 0.05$, based on a sample of over 3000 boulders. The authors also analyzed populations of smaller objects, like cobbles and pebbles, in individual high-resolution images and found these to have smaller exponents (around $-2$). This was taken as evidence for some boulders to be buried in finer particles. Ryugu's exponent for the global boulder population is significantly different from that of Vesta. Like Itokawa, Ryugu's boulders are thought to all derive from the disruption of a parent body.

{\it 101955~Bennu.} \citeA{DG19} determined the exponent for the global block distribution of B-type asteroid Bennu to be $\alpha = -2.9 \pm 0.3$, based on a sample of about 350 boulders. Bennu's exponent is significantly different from that of Vesta. Like Itokawa and Ryugu, Bennu is a suspected rubble-pile and its boulders are thought to all derive from the disruption of a parent body.

Other power law exponents for boulder distributions in the literature were derived from conventional methods like fitting the binned cumulative distribution. We must assess these exponents in light of the results of our simulations. Those derived by fitting the cumulative distribution may not be directly comparable. But those derived by fitting to a differential distribution should be comparable if the number of boulders included in the fit is large enough ($n > 100$).

{\it 4179~Toutatis.} \citeA{J15} analyzed the boulder distribution on bi-lobed, S-type asteroid Toutatis. They reported an exponent of $-4.4 \pm 0.1$ for a conventional power law fit to the binned cumulative distribution. The roll-over of their cumulative distribution due to the image resolution starts around a boulder size of 25~m, which corresponds to about 70 boulders of reliable size in the 25-61~m range, which may be too few for a reliable estimate and thus the uncertainty is almost certainly too small. The authors also considered the boulder populations of each lobe individually, and found two apparently significantly different exponents. Our simulations (Fig.~\ref{fig:generic_acuracy}) show that the derived exponents of populations of about 30-40 boulders may be very different simply by chance, both when using the ML method and fitting the differential distribution. We are unable to assess the situation for the method of fitting the binned cumulative distribution, as there is no correct way of doing that, but we expect a similar degree of inaccuracy. Therefore, the two exponents may not be significantly different in reality.

{\it 21~Lutetia.} \citeA{K12} determined an exponent of $-4$ for asteroid Lutetia based on a conventional power law fit to the differential distribution (the authors quoted a value of $-5$, but their bin size was constant on a linear rather than logarithmic scale). However, this value is essentially meaningless when considered in isolation, as their sample comprised only 6~boulders larger than 4~{\sc osiris} image pixels (240~m; compare the simulation in Fig.~\ref{fig:generic_acuracy}B).

{\it Phobos.} \citeA{T00} counted boulders on the Martian moon Phobos. They found most to be associated with Stickney crater, although this was questioned by \citeA{Ba14}, who suggested the age of Stickney to be much larger than the survival time of the boulders. \citeA{T00} obtained an exponent of $-3.2$ by fitting the cumulative distribution of all boulders they identified. We re-analyze their counts by means of the ML method. First, we note that the full data set of \citeauthor{T00} consists of the combined counts from two images of the Mars Global Surveyor camera with different spatial resolution. Images 50103 and 55103 have a resolution of 4.0 and 1.4~m per pixel, respectively. If we want to estimate a power law exponent that is representative for global Phobos, combining counts from images with different resolution will skew the SFD. We therefore assess the two images individually. When we apply our criterion for the minimum boulder size of 4~pixels, then images 50103 and 55103 have 17 and 529~boulders larger than 16 and 5.6~m, respectively. If we combine the two data sets and choose a minimum size of 16~m, then only 21~boulders satisfy this criterion. Thus, only image 55103 has a sufficiently large number of boulders to reliably retrieve the exponent. We show the SFD of boulders in image 55103 in Fig.~\ref{fig:Phobos}. The best-fit power law has an exponent of $-3.8 \pm 0.2$. However, a power law is not the correct model for the data, as confirmed by the ML test. This is not apparent in the cumulative representation (B), but can be seen in the differential representation (A), where the values in several bins are significantly off the best-fit power law curve. When we estimated $d_{\rm min}$ from the data (5~m), the derived power law exponent appeared much too small ($-2.5$), which is why we do not include this solution in Fig.~\ref{fig:Phobos}. One constraint of the Phobos data is the discrete nature of the sizes (1~m accuracy), which leads to an empty bin around 2.5~m diameter in the differential distribution (containing boulders in the 2.2-2.8~m size range) and the occurrence of steps at small diameters in the (unbinned) cumulative distribution. It may also affect the occupancy rate of bins in the differential distribution at intermediate diameters. One way to resolve this issue is to choose wider bins, but this conflicts with the image resolution of 1.4~m per pixel. We will accept the retrieved exponent for Phobos, with the caveat that a power law does not fit these data well.

{\it 21~Eros.} \citeA{T01} counted boulders on the S-type asteroid Eros and found that most are associated with Shoemaker crater. They obtained an exponent of $-3.2$ for the global boulder population by fitting the cumulative distribution. We re-analyze their counts by means of the ML method. The Eros data set combines counts from a collection of images of the NEAR-Shoemaker camera, and most boulders were measured at spatial resolutions between 2 and 5~m per pixel (99\% of the boulders were measured at a resolution $<5$~m per pixel). We adopt a minimum boulder size of 4.5~m per pixel, as 90\% of the boulders were measured at a higher resolution than that. There are 3347 boulders larger than the 4~pixels criterion (18~m). We show the boulder SFD in Fig.~\ref{fig:Eros}. The best-fit power law fits the data well, as confirmed by the ML test, and has an exponent of $-3.31 \pm 0.06$. Letting the ML algorithm estimate the minimum size, it found $d_{\rm min} = 16$~m and the same value for the exponent ($n = 4850$). Because a power law describes the data so well, the ML exponent is close to the $-3.2$ found by \citeA{T01} by fitting the cumulative distribution. Only at the largest sizes do the boulder numbers deviate from the power law curve. The authors remarked that the power law is steeper here, which we also noted for Vesta. However, as the number of boulders responsible is small ($<20$), the deviation may be due to chance. The large number of boulders ensures a small uncertainty in the exponent of the best-fit power law, and we conclude that the Eros exponent is significantly different from that of Vesta.

{\it Moon.} \citeA{C95} counted boulders at several Surveyor landing sites on the Moon as imaged by the Lunar Orbiter probe. Only the Surveyor~VII site has boulders in the size range that we consider in this paper ($> 10$~m). The authors fitted a power law to the cumulative SFD and found an exponent of $-4.0 \pm 0.1$. As they provided a table of the numbers in the size bins, we can fit a power law to the differential distribution instead (Fig.~\ref{fig:Moon}). We find an exponent of $-3.6 \pm 0.1$, for 628~boulders in the 13-80 m size range. When comparing the boulder counts in the Lunar Orbiter images (m-sized) with counts of particles seen in Surveyor images (mm to dm-sized) performed by \citeA{SM68}, \citeA{C95} noted the smaller power law exponents associated with the latter (around $-2$). The discrepancy led them to suspect that the lunar particle SFD cannot be described by a single power law over the entire size range spanning three orders of magnitude (mm to m), and that the distribution is steeper at larger sizes. However, \citeA{L17} revisited this topic and fitted power laws to the cumulative SFD of boulders (m-sized) near the Surveyor landing sites using Lunar Reconnaissance Orbiter images. The authors found exponents in the range of $-1.5$ to $-3.6$, consistent with \citeA{SM68}, and suggested that, in fact, a single power law can describe the lunar particle SFD over the entire size range from mm to m. \citeA{BM10} fitted power laws to the cumulative distribution of boulders around 18~lunar craters, and found exponents in the range of $-2.2$ to $-5.5$, with most between $-3.0$ and $-4.5$. The size range of the involved boulder populations was on the order of 1-10~m for some craters and 10-100~m for others. \citeA{K16} fitted power laws to the cumulative SFD of large numbers of boulders around Censorinus crater, grouped according to sector, and found exponents in the range of $-2.5$ to $-3.3$, with a typical boulder size range of 2-40~m. \citeA{P19} fitted a power law to the cumulative SFD of a large number of boulders around Linn\'e crater and found an exponent of $-4.0$ ($n = 12,038$, size range 4-30~m). At the large size end of the SFD, the power law predicts more boulders than were observed. This led the authors to fit a Weibull distribution, which matched the data better. Almost all published power law exponents for lunar boulders are smaller (less negative) than that for Vesta boulders, keeping in mind that they were estimated by fitting the cumulative SFD instead of ML. The $-3.6$ exponent from Fig.~\ref{fig:Moon} seems to be typical for lunar boulders in the decameter size range.

\subsubsection{Synthesis}

In the previous section we evaluated, and in some cases re-analyzed, power law exponents published for boulders on small Solar System bodies. When we adopted the power law to describe the SFD of all boulders on Vesta, we found an exponent that is significantly different from that of all other bodies, with the singular exception of Ceres. What is the physical meaning of this difference? In the literature, the steepness of the best-fit power law is often interpreted in terms of degree of fragmentation \cite{T01,K16,M19}, accompanied by a reference to \citeA{H69}, whose paper still seems to be the prime source of information on the meaning of the exponent. \citeA{H69} found that simple fragmentation results in small exponents ($-2.1$ to $-2.4$), whereas a hypervelocity impact results in a large exponent ($-3.6$), where he noted that such an impact resembles extensive regrinding. Laboratory impact experiments can provide additional insight into the meaning of the power law exponent, but are typically performed on scales orders of magnitude below that of the boulders we study. The type of experiment that is probably most relevant for planetary boulder formation involves impacts on semi-infinite surfaces. \citeA{B14} reviewed the power law exponents derived from such experiments \cite{G63,Ho69,F77,C85}. For particles up to a centimeter in size, the exponents were all around $-2.5$, regardless of the type of target material, be it sandstone, granite, basalt, or water ice. Additionally, \citeA{B14} observed no correlation between the exponent and the imparted energy density (impact kinetic energy per target mass). The authors noted that the results are different for experiments in which the entire target is disrupted, but it is unclear whether these outcomes can be extrapolated to the size range of large boulders.

While we found that the power law is not a satisfactory model for the Vesta and Ceres boulder SFDs, let us assume for the moment that it is, and compare the exponents with those derived for other small Solar System bodies. Table~\ref{tab:exponents} lists those exponents that we consider most reliable, i.e.\ preferably derived with the ML method from populations of at least 100~boulders. Figure~\ref{fig:power_law_exponents} displays the exponents as a function of boulder size, indicating for each the size range from which it was obtained. We distinguish between boulders formed by impact on a surface and boulders that make up the rubble-pile asteroids Itokawa, Ryugu, and Bennu. Experimental evidence suggests that the SFD resulting from fragmentation of a target body is different from that resulting from an impact on a semi-infinite surface \cite{B14}. But as it is not clear whether these results can be extrapolated to planetary scales, we include rubble-pile bodies in the figure. There are two data points for both Vesta and Ceres: one for a minimum boulder size of $d_{\rm min} = 4$~image pixels (black symbols) and one for $d_{\rm min}$ estimated by the ML algorithm (red symbols). The boulders counted on Vesta and Ceres are large ($> 100$~m), and their exponents cluster around $-5$. Given their very different surface composition, the similarity of the exponents is surprising. Boulders identified on other bodies are smaller (10-100~m), and their exponents are smaller too, ranging from around $-3$ to $-4$. The exponents of the rubble-pile asteroids (Bennu, Itokawa, Ryugu) are not clearly separated from the others (Eros, Moon, Phobos). Figure~\ref{fig:power_law_exponents} also includes several exponents for Ryugu particles at small scales ($< 1$~m), determined from individual images \cite{M19}. These exponents are smaller yet, clustering around $-2$, consistent with exponents derived from laboratory impact experiments \cite{B14} and for the lunar regolith \cite{SM68}. The figure shows a correlation between the exponent and the particle (boulder) size range from which it is derived.

Following \citeA{H69}, the large exponents of the SFDs of Vesta and Ceres boulders imply an extremely high degree of fragmentation. But this is difficult to understand in light of the relatively low average impact velocities expected on their surface \cite{B15}. The idea that extensive regrinding experienced in a hypervelocity impact leads to a power law SFD with a large exponent \cite{H69} is not supported by grinding experiments, which typically result in a Weibull distribution \cite{RR33,MM77,DS13}. Furthermore, \citeA{B89} and \citeA{BW95} theorized that a power law distribution results from a single fragmentation event, whereas sequential fragmentation (i.e., regrinding) results in a Weibull distribution. We found that the power law is not a good model for the Vesta and Ceres boulder SFDs. However, the Weibull distribution satisfactorily fits the data. This suggests that the unusually large exponents for Vesta and Ceres should not be interpreted in terms of degree of fragmentation, but simply follow from the shape of the Weibull distribution. The dependence of the power law exponent on particle size range in Fig.~\ref{fig:power_law_exponents} can be understood if, in general, the particle SFD on small bodies follows a Weibull distribution over a wide range of sizes. This would provide a natural explanation for the relatively low abundance of larger boulders, which are sometimes reported as ``missing'' \cite{T01,M19}. The Weibull distribution has been noted before to fit the particle SFD, both at small \cite{MM77} and large scales \cite{P19}. Over a narrow size range the Weibull SFD may masquerade as a power law. The exponent of such a ``local'' power law would primarily be a function of the particle size range from which it was derived, but other factors (composition, impact velocity) may yet play a role.

\section{Conclusions}

We identified, counted, and measured more than 10,000 boulders on the surface of Vesta, with sizes up to several hundred meters. We found all boulders to be associated with impact craters. There is little evidence for boulder production by physical weathering of crater walls, so the vast majority of boulders were created upon impact. Craters with boulders are distributed mostly uniformly over the surface, only seeming to avoid a large area of below average albedo. This area is believed to be rich in carbon, delivered by the primitive impactor that created the ancient Veneneia basin \cite{R12,J14}. It is unclear why boulders are rare here; the regolith may be thicker than average \cite{D16} or the boulders live shorter. Using published crater ages, we established that Vesta boulders have a lifetime of about a few hundred million years. This time is on the same order as that estimated for meter-sized boulders on the Moon by \citeA{B15}, who predicted that such boulders live 30 times shorter on Vesta than on the Moon. One reason for the apparent disagreement may be that the Vesta boulders in our sample are an order of magnitude larger than the lunar boulders considered by the authors.

In the literature, the SFD of planetary boulders is often fitted with a power law. Different methods to derive the power law exponent (slope) are used, but only the maximum likelihood (ML) method is statistically sound \cite{C09}. We investigate how the number of boulders in a population affects the derived exponent, and confirm that the ML method is biased at low numbers. The exponent is most reliably derived for a population size of at least 100 boulders, where we recommend adopting a minimum boulder size of 4 image pixels. We derived the power law exponent for all Vesta craters with boulders, and find that the SFD of Marcia crater stands out as different from all others. The exponent for Vesta's global boulder population is around $-5$. We reviewed published exponents for small Solar System bodies. The use of different fitting techniques and limited awareness of the unreliability of small-number statistics hinder a direct comparison, and we re-analyzed several data sets. The Vesta power law slope is steeper than typically found for other small bodies. A statistical test reveals that the power law is actually not a good model for the Vesta SFD, but the Weibull distribution fits the data very well. The Weibull distribution is commonly applied to describe SFDs resulting from rock grinding experiments, and results from the fractal nature of the cracks propagating in the rock interior \cite{BW95}. The Weibull distribution may provide a better description of the SFD of boulders on small bodies than the power law, and would naturally result in a steeper SFD for the relatively large boulders of Vesta.

\acknowledgments
We are grateful for technical support provided by J-Vesta developer Dale Noss and his team at ASU. We thank Peter Thomas for kindly sharing the Eros and Phobos boulder data. We also thank Maurizio Pajola for his helpful review and Alexander Basilevsky for fruitful discussions. Dawn framing camera images are available from NASA's Planetary Data System at \url{https://pds.nasa.gov/}. Our Vesta boulder data are available at DOI:10.5281/zenodo.3833759.

\bibliography{Boulders}

\newpage
\clearpage

\begin{table}
\centering
\caption{All craters on Vesta with at least one boulder larger than 4~pixels ($d > 80$~m). Crater and boulder diameters are $D$ and $d$, respectively, and $\alpha$ is the power law exponent of the (cumulative) boulder SFD as derived with the ML method (only for craters with $n_{d > 4 {\rm px}} > 5$).}
\label{tab:craters}
\begin{tabular}{lllllllll}
\hline
Name & Longitude & Latitude & $D$ & $d_{\rm max}$ & $n_{d > 3 {\rm px}}$ & $n_{d > 4 {\rm px}}$ & $\alpha$ & $\sigma_\alpha$ \\
 & ($^\circ$E) & ($^\circ$) & (km) & (m) & & & & \\
\hline
Aelia & 140.7 & $-14.2$ & 4.7 & 94 & 22 & 4 & & \\
Angioletta & 29.3 & $-40.1$ & 18.6 & 129 & 87 & 14 & $-6.8$ & 1.8 \\
Antonia & 200.9 & $-58.9$ & 17.3 & 154 & 491 & 187 & $-5.4$ & 0.4 \\
Aquilia & 41.1 & $-50.0$ & 33.8 & 250 & 143 & 71 & $-4.0$ & 0.5 \\
Arruntia & 71.6 & $+39.4$ & 10.4 & 109 & 60 & 16 & $-8.3$ & 2.1 \\
Canuleia & 294.5 & $-33.5$ & 11.2 & 121 & 67 & 11 & $-5.8$ & 1.8 \\
Charito & 300.5 & $-44.5$ & 6.8 & 109 & 43 & 9 & $-5.3$ & 1.8 \\
Cornelia & 225.5 & $-9.0$ & 16.7 & 157 & 221 & 60 & $-5.0$ & 0.6 \\
Drusilla & 261.2 & $-14.8$ & 20.9 & 136 & 85 & 33 & $-8.5$ & 1.5 \\
Eusebia & 204.8 & $-43.1$ & 23.3 & 113 & 133 & 27 & $-6.5$ & 1.3 \\
Fabia & 265.8 & $+15.6$ & 11.9 & 152 & 259 & 86 & $-5.8$ & 0.6 \\
Fausta & 309.7 & $-25.5$ & 3.2 & 111 & 17 & 3 & & \\
Fonteia & 141.5 & $-53.5$ & 21.1 & 166 & 162 & 56 & $-4.3$ & 0.6 \\
Galeria & 228.3 & $-29.5$ & 22.9 & 170 & 69 & 31 & $-3.4$ & 0.6 \\
Gegania & 60.8 & $+3.9$ & 23.8 & 269 & 30 & 14 & $-4.0$ & 1.1 \\
Hortensia & 15.1 & $-46.6$ & 29.5 & 218 & 34 & 16 & $-5.7$ & 1.4 \\
Justina & 318.0 & $-34.0$ & 7.1 & 193 & 96 & 23 & $-5.1$ & 1.1 \\
Lepida & 306.7 & $+18.3$ & 40.2 & 146 & 30 & 11 & $-4.7$ & 1.4 \\
Licinia & 17.2 & $+23.6$ & 23.6 & 216 & 280 & 111 & $-4.1$ & 0.4 \\
Mamilia & 291.8 & $+48.3$ & 35.0 & 142 & 38 & 22 & $-5.5$ & 1.2 \\
Marcia & 190.2 & $+9.5$ & 61.0 & 390 & 958 & 476 & $-3.6$ & 0.2 \\
Numisia & 247.5 & $-7.7$ & 30.0 & 174 & 136 & 36 & $-4.4$ & 0.7 \\
Oppia & 308.9 & $-7.6$ & 34.0 & 164 & 65 & 32 & $-4.3$ & 0.8 \\
Paculla & 1.8 & $-64.0$ & 19.4 & 172 & 177 & 59 & $-5.4$ & 0.7 \\
Pinaria & 32.0 & $-29.2$ & 38.0 & 277 & 315 & 143 & $-4.4$ & 0.4 \\
Portia & 41.5 & $+0.9$ & 10.9 & 155 & 17 & 6 & $-3.5$ & 1.4 \\
Publicia & 84.4 & $+14.6$ & 16.6 & 110 & 28 & 6 & $-7.9$ & 3.2 \\
Rubria & 18.3 & $-7.4$ & 10.3 & 113 & 97 & 19 & $-7.5$ & 1.7 \\
Rufillia & 138.7 & $-13.0$ & 15.5 & 109 & 25 & 7 & $-6.0$ & 2.3 \\
Scantia & 274.6 & $+29.7$ & 16.4 & 185 & 182 & 86 & $-4.2$ & 0.5 \\
Serena & 120.6 & $-20.4$ & 19.0 & 119 & 16 & 4 & & \\
Severina & 122.7 & $-76.3$ & 33.4 & 248 & 182 & 82 & $-3.8$ & 0.4 \\
Sextilia & 146.1 & $-39.0$ & 19.8 & 234 & 69 & 25 & $-4.6$ & 0.9 \\
Sossia & 286.0 & $-37.0$ & 7.4 & 109 & 32 & 5 & & \\
Teia & 271.0 & $-3.5$ & 6.5 & 144 & 37 & 10 & $-4.7$ & 1.5 \\
Tuccia & 198.1 & $-39.7$ & 3.3 & 97 & 52 & 10 & $-10.2$ & 3.2 \\
Unnamed2 & 74.0 & $+27.5$ & 11.7 & 145 & 69 & 25 & $-4.8$ & 1.0 \\
Unnamed3 & 211.0 & $-24.0$ & 11.7 & 120 & 97 & 22 & $-7.8$ & 1.7 \\
Unnamed4 & 280.2 & $+7.9$ & 16.1 & 165 & 302 & 92 & $-5.5$ & 0.6 \\
Unnamed5 & 351.0 & $+17.6$ & 3.9 & 93 & 47 & 8 & $-15.5$ & 5.5 \\
Unnamed6 & 348.0 & $-35.0$ & 7.4 & 138 & 31 & 7 & $-4.0$ & 1.5 \\
Unnamed7 & 175.8 & $+33.2$ & 22.5 & 104 & 19 & 6 & $-7.5$ & 3.0 \\
Unnamed8 & 116.0 & $+32.7$ & 20.5 & 169 & 41 & 17 & $-5.6$ & 1.4 \\
Unnamed9 & 297.8 & $+37.6$ & 10.4 & 130 & 16 & 3 & & \\
Unnamed10 & 358.4 & $+15.1$ & 10.1 & 107 & 66 & 13 & $-10.6$ & 2.9 \\
\hline
\end{tabular}
\end{table}
\begin{table}
\centering
\begin{tabular}{lllllllll}
\hline
Name & Longitude & Latitude & $D$ & $d_{\rm max}$ & $n_{d > 3 {\rm px}}$ & $n_{d > 4 {\rm px}}$ & $\alpha$ & $\sigma_\alpha$ \\
 & ($^\circ$E) & ($^\circ$) & (km) & (m) & & & & \\
\hline
Unnamed11 & 269.3 & $-73.1$ & 6.2 & 97 & 20 & 7 & $-12.8$ & 4.8 \\
Unnamed12 & 291.8 & $-73.8$ & 7.5 & 217 & 99 & 32 & $-4.4$ & 0.8 \\
Unnamed13 & 245.2 & $-50.3$ & 5.8 & 98 & 50 & 11 & $-11.9$ & 3.6 \\
Unnamed14 & 307.1 & $-1.9$ & 5.1 & 142 & 44 & 4 & & \\
Unnamed15 & 44.8 & $+53.8$ & 11.7 & 90 & 19 & 5 & & \\
Unnamed17 & 340.3 & $+32.5$ & 9.7 & 90 & 11 & 4 & & \\
Unnamed18 & 225.2 & $-39.9$ & 4.3 & 81 & 19 & 1 & & \\
Unnamed19 & 22.7 & $+8.8$ & 11.3 & 106 & 23 & 5 & & \\
Unnamed20 & 39.9 & $+23.6$ & 7.5 & 83 & 12 & 2 & & \\
Unnamed21 & 129.0 & $-49.2$ & 12.8 & 120 & 20 & 8 & $-5.3$ & 1.9 \\
Unnamed22 & 72.4 & $+11.6$ & 5.2 & 85 & 7 & 3 & & \\
Unnamed23 & 263.1 & $+4.6$ & 9.3 & 115 & 17 & 4 & & \\
Unnamed24 & 329.8 & $+0.4$ & 9.6 & 122 & 23 & 7 & $-5.5$ & 2.1 \\
Unnamed26 & 340.8 & $+27.4$ & 4.8 & 88 & 7 & 2 & & \\
Unnamed27 & 18.1 & $-14.9$ & 7.3 & 94 & 26 & 2 & & \\
Unnamed28 & 85.4 & $-22.5$ & 7.6 & 103 & 15 & 6 & $-5.5$ & 2.3 \\
Unnamed30 & 281.4 & $-31.8$ & 7.3 & 88 & 25 & 3 & & \\
Unnamed32 & 331.7 & $-6.2$ & 7.9 & 120 & 37 & 12 & $-8.7$ & 2.5 \\
Unnamed33 & 12.7 & $-38.0$ & 4.4 & 90 & 12 & 2 & & \\
Unnamed34 & 159.4 & $-42.4$ & 8.8 & 166 & 29 & 12 & $-4.1$ & 1.2 \\
Unnamed35 & 309.2 & $-48.3$ & 7.9 & 105 & 45 & 9 & $-8.5$ & 2.8 \\
Unnamed36 & 300.4 & $-55.6$ & 10.4 & 199 & 116 & 55 & $-5.1$ & 0.7 \\
Unnamed37 & 285.1 & $-56.3$ & 7.2 & 131 & 88 & 27 & $-8.1$ & 1.6 \\
Vibidia & 220.5 & $-27.0$ & 7.1 & 163 & 300 & 91 & $-6.6$ & 0.7 \\
\hline
\end{tabular}
\end{table}

\begin{table}
\centering
\caption{Age and boulder density for craters for which an age estimate is available. Density is defined as the number of boulders larger than 3~pixels divided by crater equivalent area. The densities in brackets are underestimates, as the associated craters were largely in the shadow in {\sc lamo} images.}
\begin{tabular}{llll}
\hline
Name & Age & Density$^a$ & Source for age \\
     & (Ma) & (km$^{-2}$) & \\
\hline
Antonia & 19-23 & $2.1 \pm 0.1$ & \citeA{SK14} \\
Arruntia & 2-3 & ($0.71 \pm 0.09$) & \citeA{R14} \\
Cornelia & 9-14 & $1.01 \pm 0.07$ & \citeA{Kr14} \\
Eusebia & 208-221 & $0.31 \pm 0.03$ & \citeA{K14} \\
Galeria & 209-241 & $0.17 \pm 0.02$ & \citeA{K14} \\
Licinia & 45-54 & $0.64 \pm 0.04$ & \citeA{R14} \\
Mamilia & 164-188 & ($0.039 \pm 0.006$) & \citeA{R14} \\
Marcia & 120-149 & $0.33 \pm 0.01$ & \citeA{W14a} \\
Oppia & 309-331 & $0.072 \pm 0.009$ & \citeA{SK14} \\
Rubria & 14-23 & $1.16 \pm 0.12$ & \citeA{Kr14} \\
Scantia & 129-149 & $0.86 \pm 0.06$ & \citeA{R14} \\
Unnamed36 & 240-270 & $1.37 \pm 0.13$ & \citeA{Kr14} \\
Vibidia & 9-11 & $7.6 \pm 0.4$ & \citeA{K14} \\
\hline
\end{tabular}
\label{tab:ages}
\end{table}

\begin{table}
\centering
\caption{Power law exponents determined from populations of at least 100~boulders on small Solar System bodies. Brackets around an exponent indicate that a power law is not a good model for the data as indicated by the ML test (unavailable for the Moon).}
\begin{tabular}{llll}
\hline
Body & Exponent & Method  & Source \\
\hline
Bennu & $-2.9 \pm 0.3$ & ML ($d_{\rm min}$ estimated) & \citeA{DG19} \\
Ceres & ($-4.3 \pm 0.2$) & ML ($d_{\rm min} = 4$ px) & This work, based on \citeA{S18} \\
      & ($-5.6 \pm 0.4$) & ML ($d_{\rm min}$ estimated) & ibid. \\
Eros & $-3.31 \pm 0.06$ & ML ($d_{\rm min} = 4$ px) & This work, based on \citeA{T01} \\
     & $-3.31 \pm 0.06$ & ML ($d_{\rm min}$ estimated) & ibid. \\
Itokawa & $-3.6 \pm 0.3$ & ML ($d_{\rm min}$ estimated) & \citeA{DS15} \\
        & $-3.5 \pm 0.2$ & ML ($d_{\rm min}$ estimated) & \citeA{M19} \\
Moon & $-3.6 \pm 0.1$ & Fitting differential & This work, based on \citeA{C95} \\
Phobos & ($-3.8 \pm 0.2$) & ML ($d_{\rm min} = 4$ px) & This work, based on \citeA{T00} \\
Ryugu & $-2.65 \pm 0.05$ & ML ($d_{\rm min}$ estimated) & \citeA{M19} (global) \\
      & $-2.07 \pm 0.06$ & ML ($d_{\rm min}$ estimated) & ibid.\ (local) \\
      & $-2.01 \pm 0.06$ & ML ($d_{\rm min}$ estimated) & ibid.\ (local) \\
      & $-1.96 \pm 0.07$ & ML ($d_{\rm min}$ estimated) & ibid.\ (local) \\
      & $-1.98 \pm 0.09$ & ML ($d_{\rm min}$ estimated) & ibid.\ (local) \\
      & $-1.65 \pm 0.05$ & ML ($d_{\rm min}$ estimated) & ibid.\ (local) \\
Vesta & ($-4.7 \pm 0.1$) & ML ($d_{\rm min} = 4$ px) & This work \\
      & ($-4.8 \pm 0.1$) & ML ($d_{\rm min}$ estimated) & ibid. \\
\hline
\end{tabular}
\label{tab:exponents}
\end{table}

\newpage
\clearpage

\begin{figure}
\centering
\includegraphics[width=8cm,angle=0]{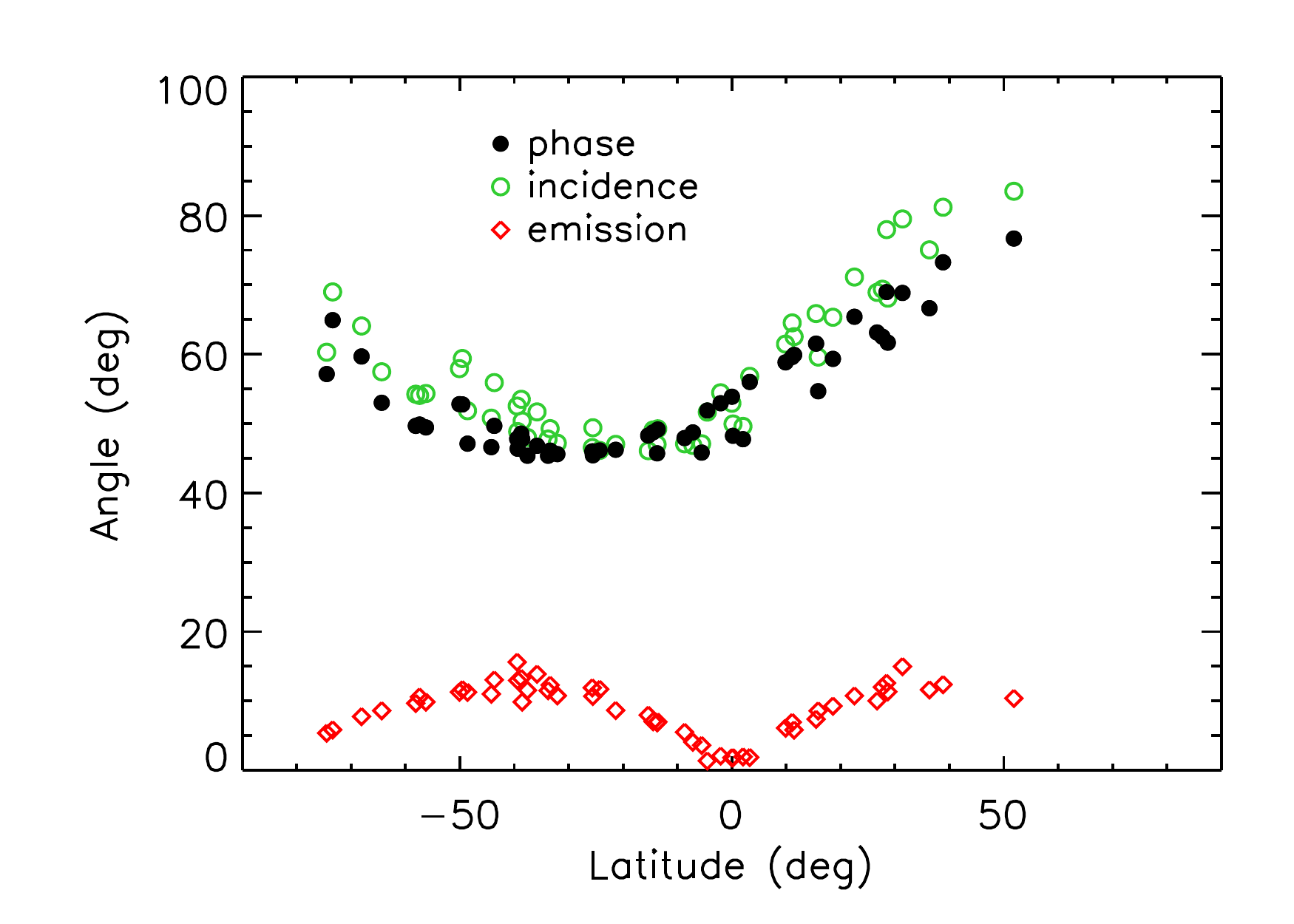}
\caption{Boulder viewing conditions: Photometric angles at the center of selected {\sc lamo} images.}
\label{fig:viewing_angles}
\end{figure}

\begin{figure}
\centering
\includegraphics[width=\textwidth,angle=0]{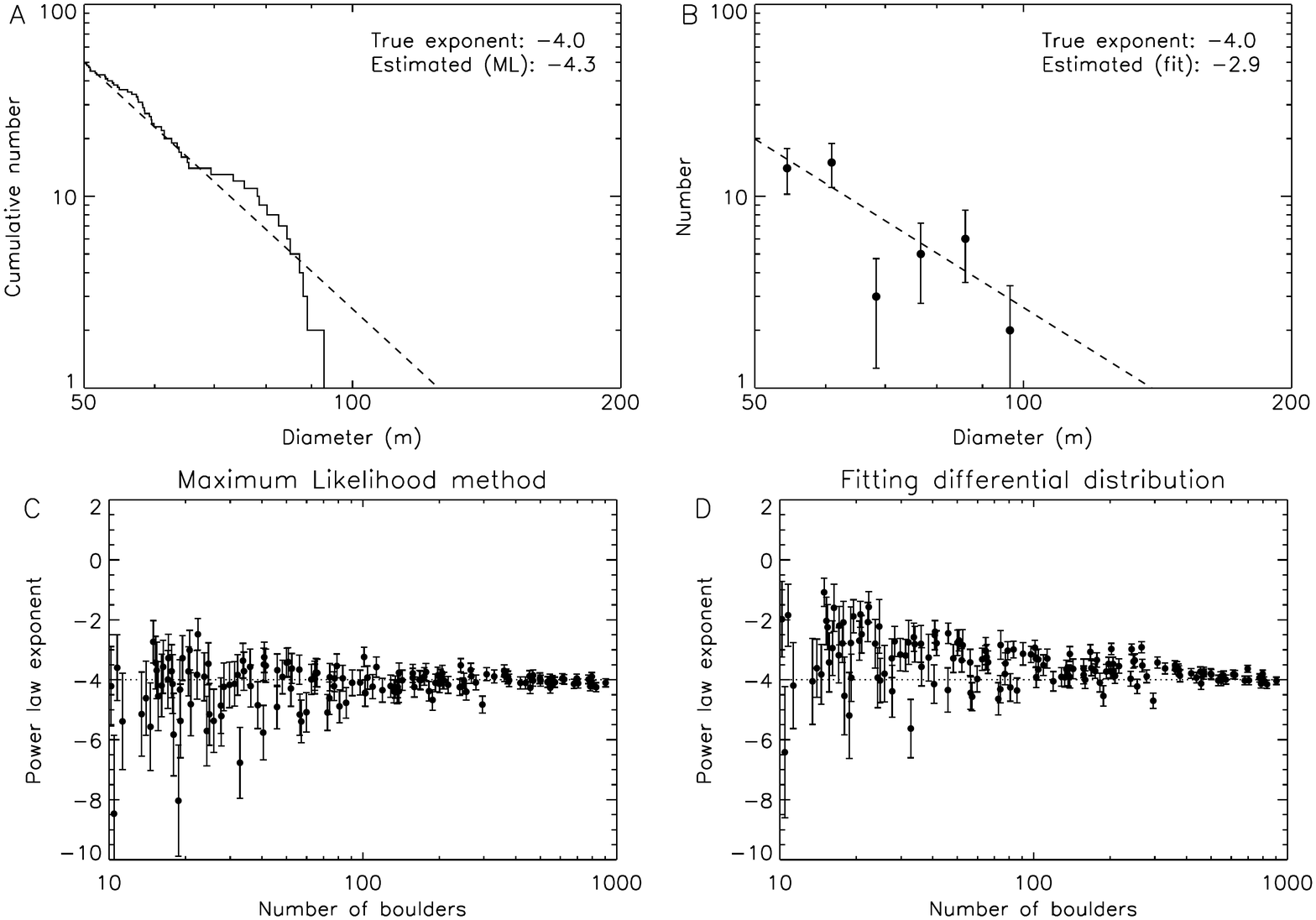}
\caption{Monte Carlo simulations of boulder populations, generated assuming the cumulative SFD follows a power law with exponent $-4$. {\bf A} \& {\bf B}: A population of 50 boulders with a minimum size of 50~m, shown as (A) a cumulative plot with the power law exponent estimated by the ML method and (B) a binned, differential plot with the exponent estimated by a least-squares fit. The dashed lines are the best-fit power laws. {\bf C} \& {\bf D}: Estimating the power law exponents of 150 boulder populations by (C) ML and (D) fitting the differential distribution. The populations were randomly generated in the logarithmic interval $(10, 1000)$ for the number of boulders. The dotted line indicates the exponent's true value.}
\label{fig:generic_acuracy}
\end{figure}

\begin{figure}
\centering
\includegraphics[width=10cm,angle=0]{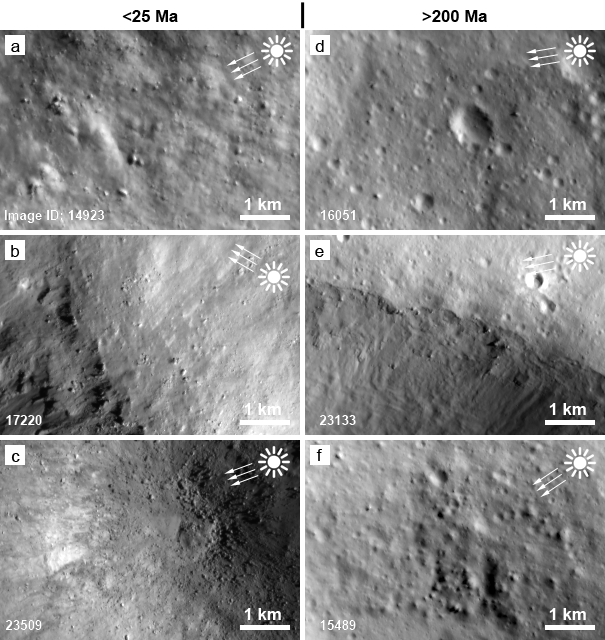}
\caption{Examples of boulder fields associated with craters younger than 25~Ma (left side) and older than 200~Ma (right side) (ages in Table~\ref{tab:ages}). North is up in all images. The illumination direction is shown at the upper right and the image number at lower left. {\bf a}.~Southern (downslope) inner wall of Antonia. {\bf b}.~NE rim of Cornelia and adjacent plateau. {\bf c}.~Interior of Vibidia. {\bf d}.~Southern part of Eusebia. {\bf e}.~Northern rim of Galeria. {\bf f}.~Northern interior wall of Unnamed36.}
\label{fig:boulders}
\end{figure}

\begin{figure}
\centering
\includegraphics[width=\textwidth,angle=0]{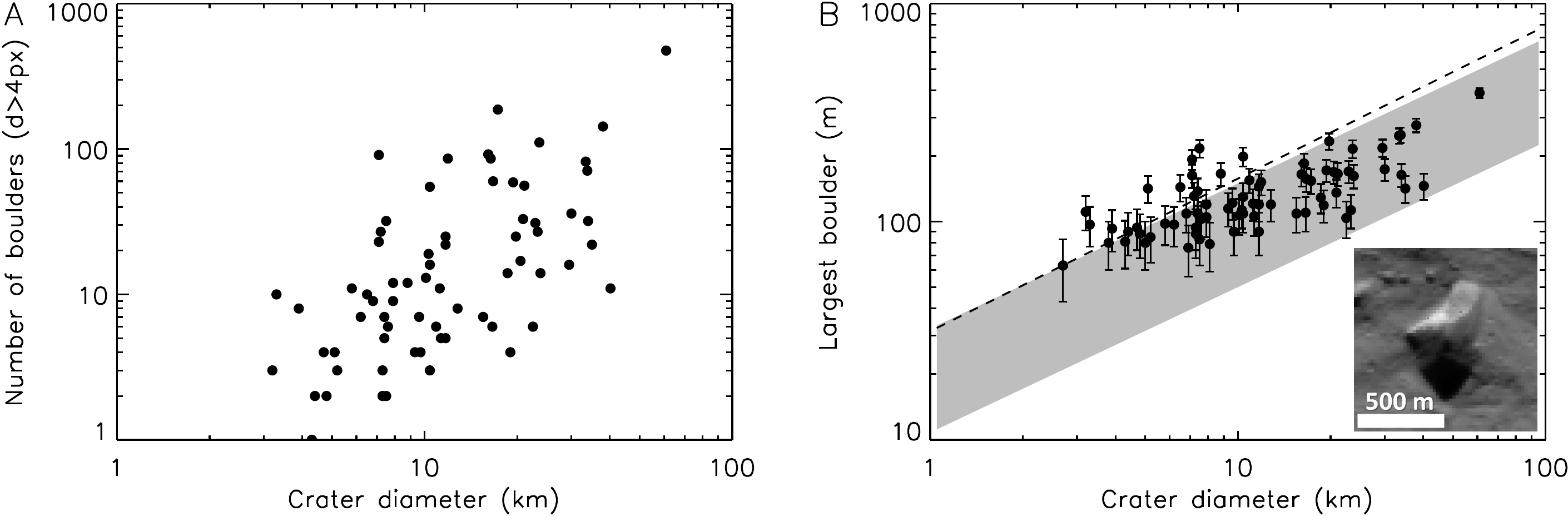}
\caption{Basic statistics of the Vesta boulder population. {\bf A}.~Number of boulders larger than 4~image pixels ($d > 80$~m) for all craters. {\bf B}.~Diameter of largest boulder for these craters, assuming a measurement error of 1~pixel. The empirical range given by \citeA{M71} for selected lunar and terrestrial craters is shown in gray. The dashed line is the relation given by \citeA{L96}. The inset shows the largest boulder identified on Vesta, located on the floor of Marcia crater.}
\label{fig:basic_stats}
\end{figure}

\begin{figure}
\centering
\includegraphics[width=\textwidth,angle=0]{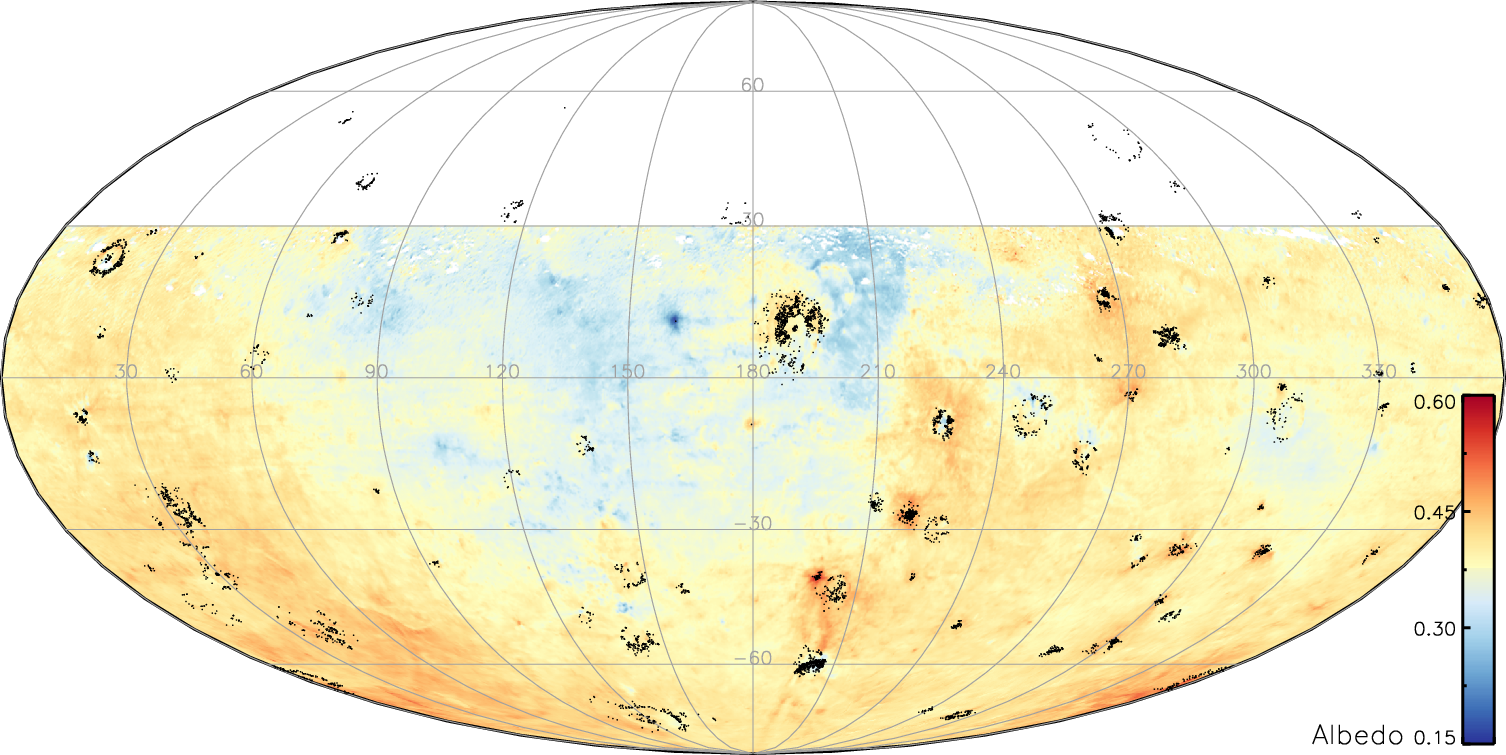}
\caption{Distribution of all boulders (black dots) identified on Vesta with a size of at least 3~pixels ($d > 60$~m) displayed on a normal albedo map from \citeA{S13a}.}
\label{fig:global}
\end{figure}

\begin{figure}
\centering
\includegraphics[width=\textwidth,angle=0]{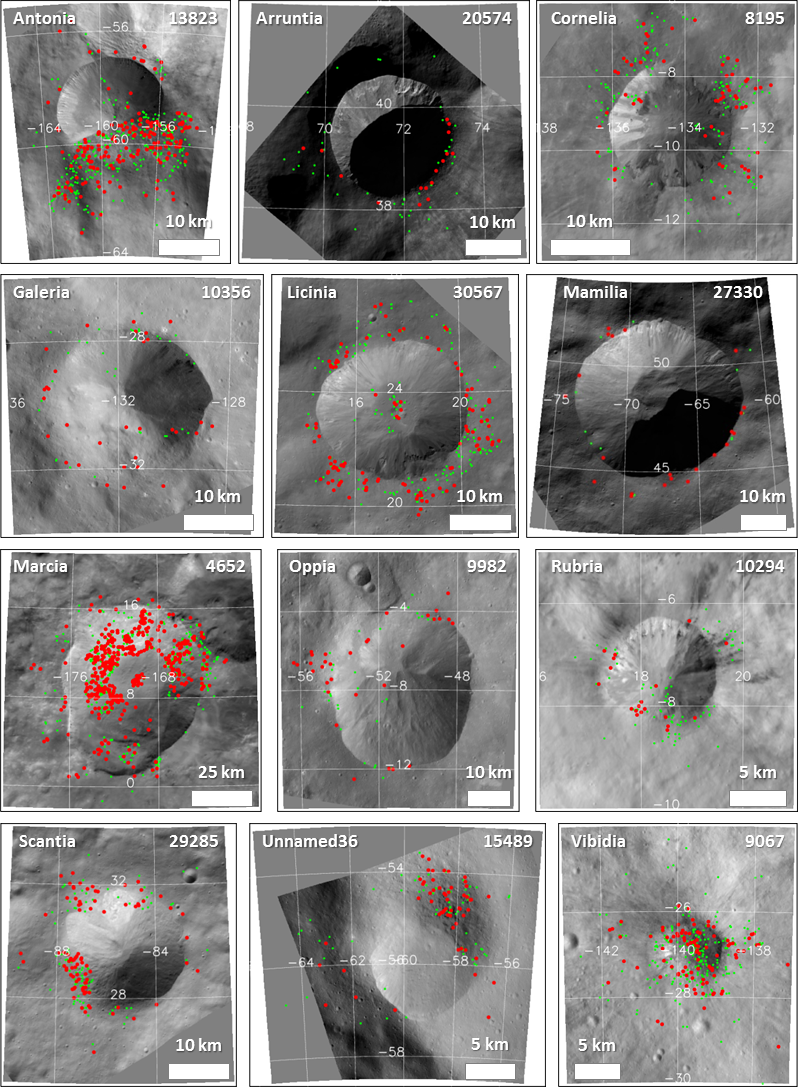}
\caption{Boulders around craters for which an age estimate is available (Table~\ref{tab:ages}). Green, small dots represent boulders with a size between 3 and 4~pixels ($60$~m~$< d < 80$~m). Red, large dots represent boulders larger than 4~pixels ($d > 80$~m). The image number is indicated in the top right.}
\label{fig:projections}
\end{figure}

\begin{figure}
\centering
\includegraphics[width=8cm,angle=0]{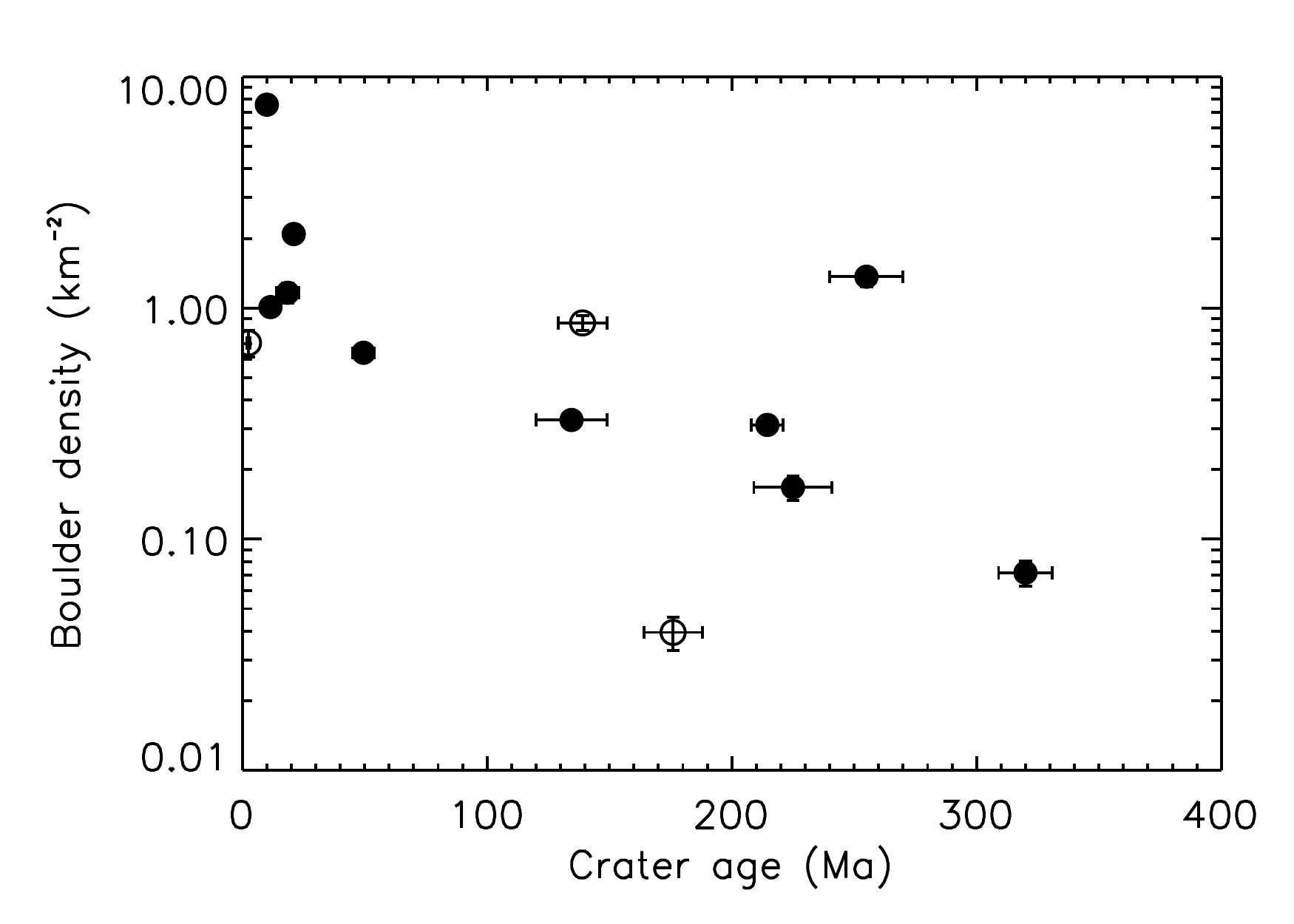}
\caption{The density of boulders larger than 3~pixels ($d > 60$~m) versus crater age (Table~\ref{tab:ages}). The error bars on the density were calculated assuming the number of boulders follows a Poisson distribution. The open symbols represent craters whose boulder density is underestimated because they were largely in the shadow in {\sc lamo} images.}
\label{fig:density_vs_age}
\end{figure}

\begin{figure}
\centering
\includegraphics[width=\textwidth,angle=0]{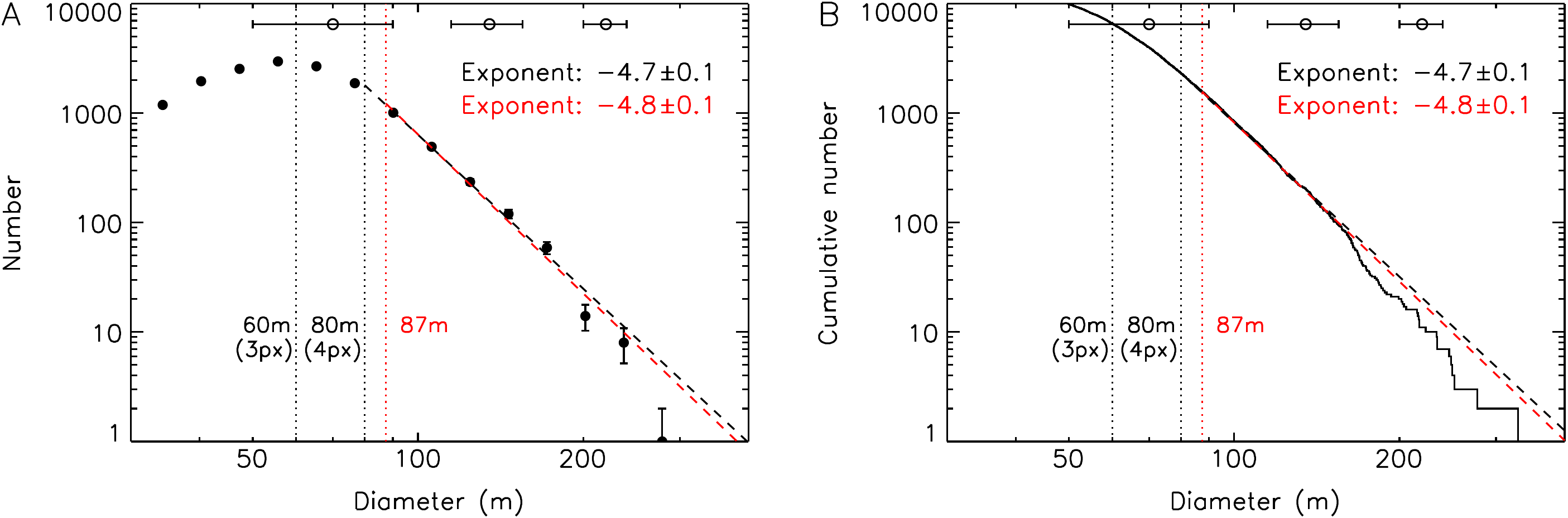}
\caption{The SFD of all boulders identified on Vesta, displayed both in differential ({\bf A}) and cumulative ({\bf B}) format. Different size limits are indicated by vertical (dotted) lines. The dashed lines are best-fit power laws using the ML method, with exponent indicated: The black dashed line has $d_{\rm min} = 80$~m (4~pixels), whereas the red dashed line has $d_{\rm min} = 87$~m, as estimated by the ML algorithm. The error bars at the top indicate the uncertainty in boulder size at different diameters due to a 1~pixel measurement error.}
\label{fig:global_ML}
\end{figure}

\begin{figure}
\centering
\includegraphics[width=\textwidth,angle=0]{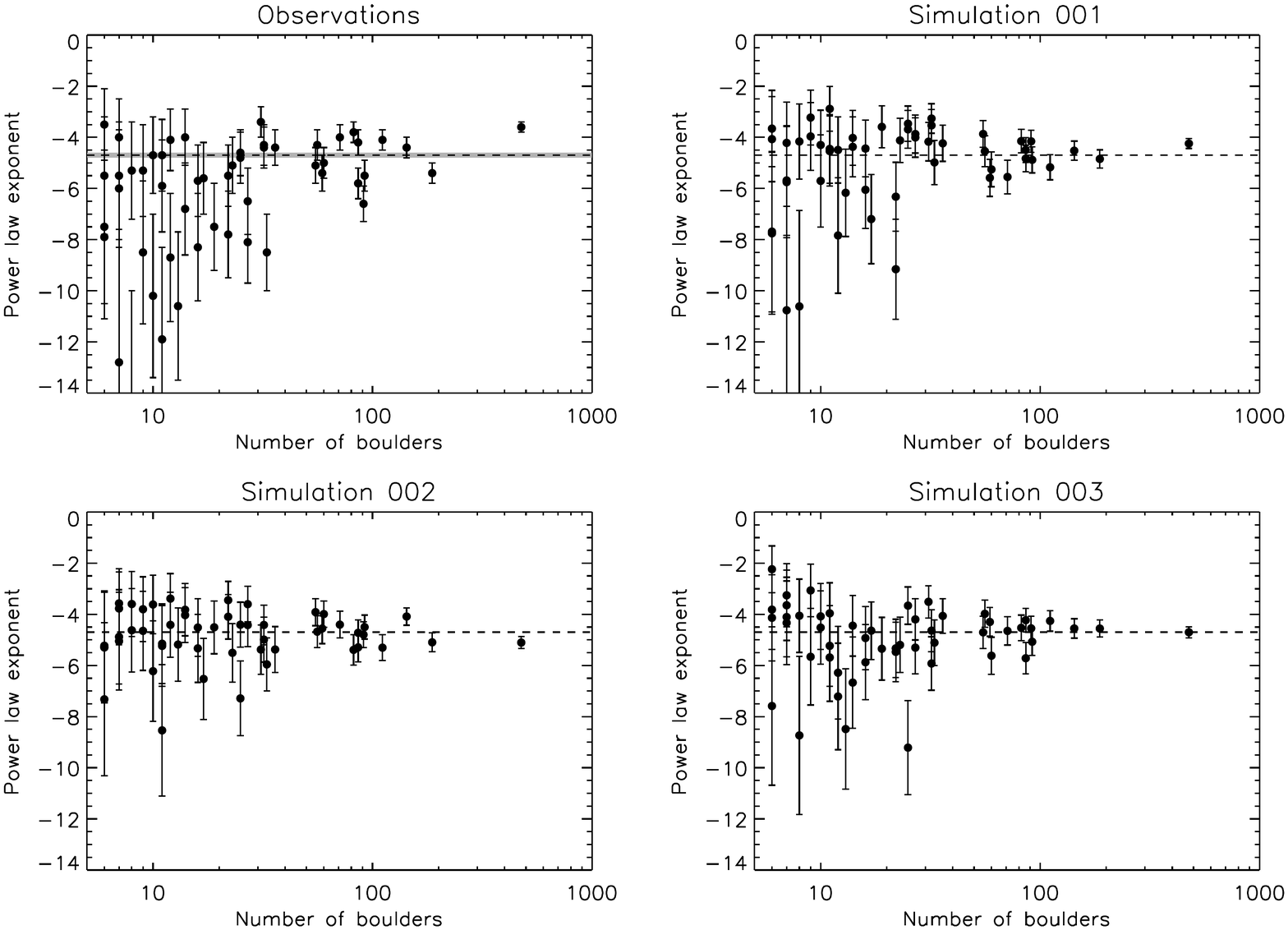}
\caption{Power law exponents for all craters with a population of at least 6 boulders larger than 4~pixels ($n = 52$). The observed exponents were derived by fitting a power law to the data of each crater. The best fit power law index for the observed global boulder distribution is $\alpha = -4.7 \pm 0.1$ (dashed line with gray confidence interval). The crater with the largest number of boulders (548) is Marcia. We compare the observations to three simulations. The simulated exponents were derived by fitting randomly generated boulder distributions, assuming a Pareto distribution with $\alpha = -4.7$ (dashed line), using the number of boulders in the population of each crater as input.}
\label{fig:accuracy}
\end{figure}

\begin{figure}
\centering
\includegraphics[width=\textwidth,angle=0]{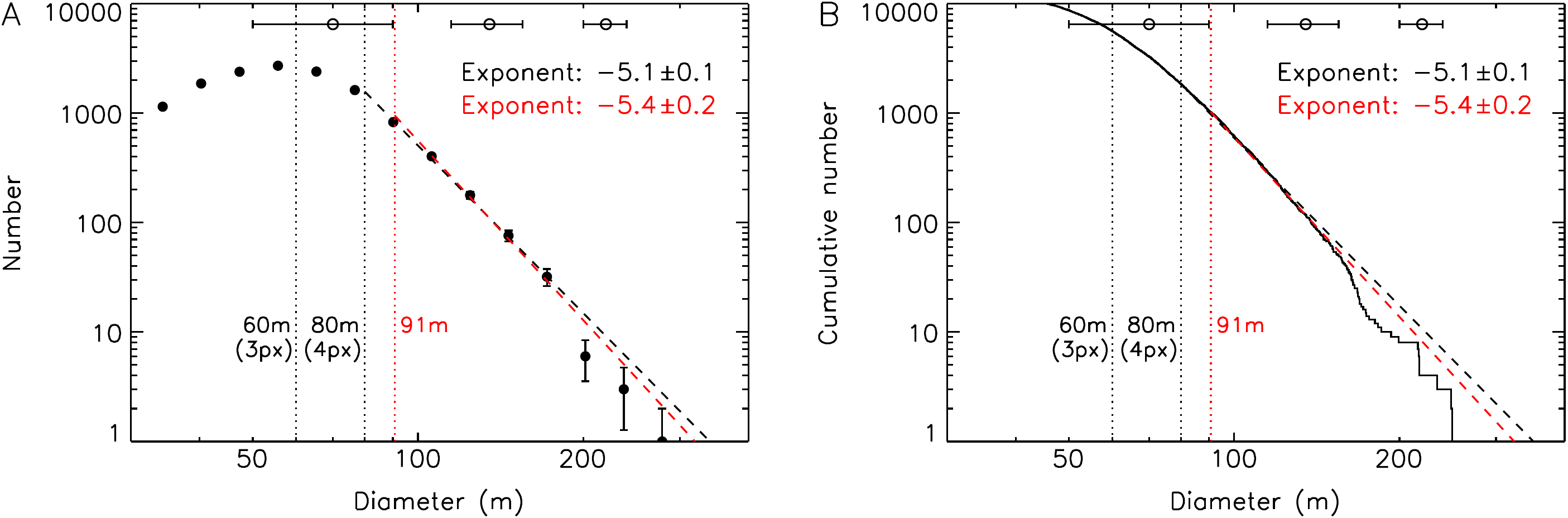}
\caption{The SFD of all boulders identified on Vesta, excluding those of Marcia, displayed both in differential ({\bf A}) and cumulative ({\bf B}) format. Different size limits are indicated by vertical (dotted) lines. The dashed lines are best-fit power laws using the ML method, with exponent indicated: The black dashed line has $d_{\rm min} = 80$~m (4~pixels), whereas the red dashed line has $d_{\rm min} = 91$~m, as estimated by the ML algorithm. The error bars at the top indicate the uncertainty in boulder size at different diameters due to a 1~pixel measurement error.}
\label{fig:global_ML_noMarcia}
\end{figure}

\begin{figure}
\centering
\includegraphics[width=\textwidth,angle=0]{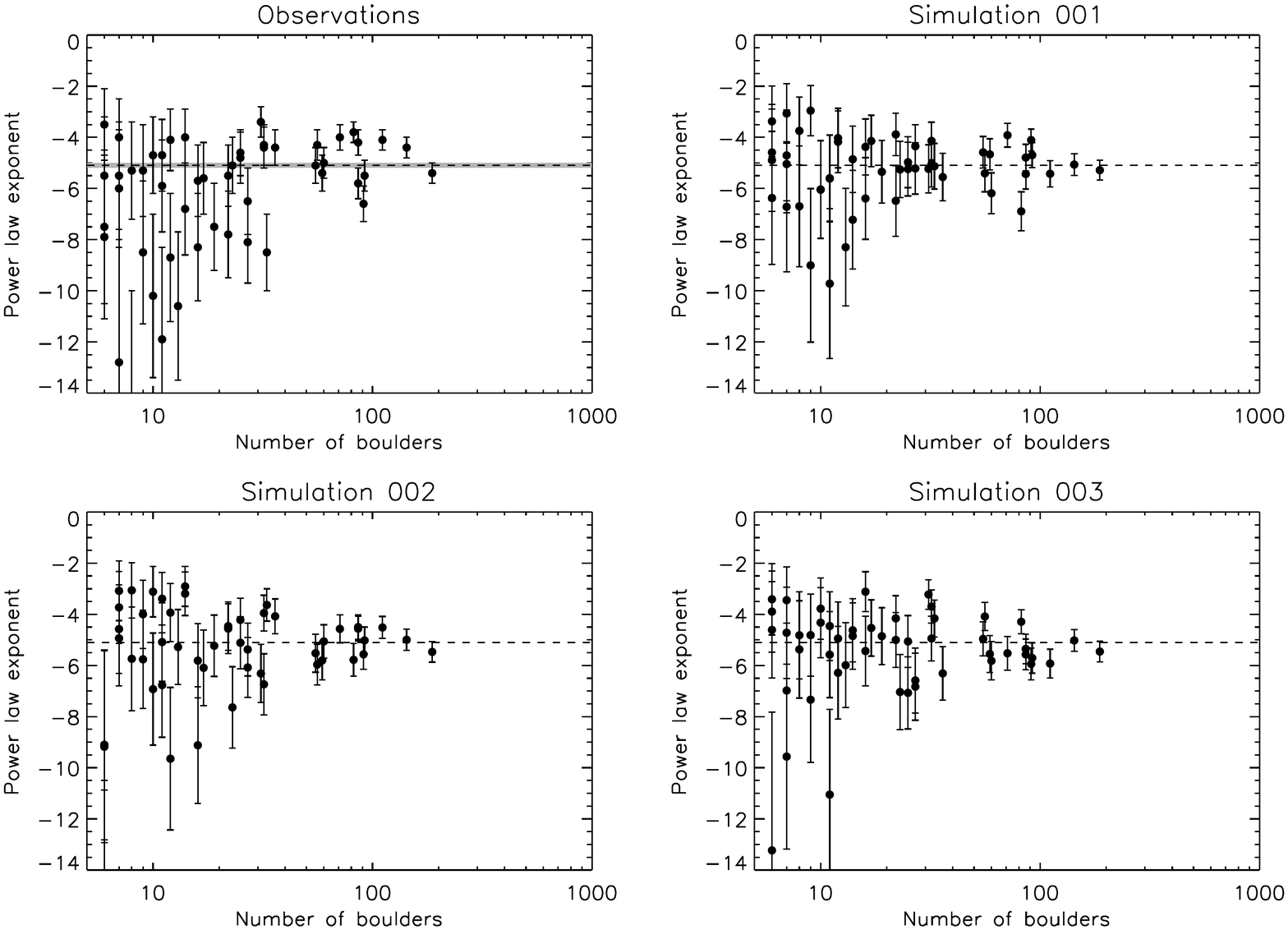}
\caption{Power law exponents for all craters with a population of at least 6 boulders larger than 4~pixels, excluding Marcia crater ($n = 51$). The observed exponents were derived by fitting a power law to the data of each crater. The best fit power law index for the observed global boulder distribution minus Marcia is $\alpha = -5.1 \pm 0.1$ (dashed line with gray confidence interval). We compare the observations to three simulations. The simulated exponents were derived by fitting randomly generated boulder distributions, assuming a Pareto distribution with $\alpha = -5.1$ (dashed line), using the number of boulders in the population of each crater as input.}
\label{fig:accuracy_noMarcia}
\end{figure}

\begin{figure}
\centering
\includegraphics[width=\textwidth,angle=0]{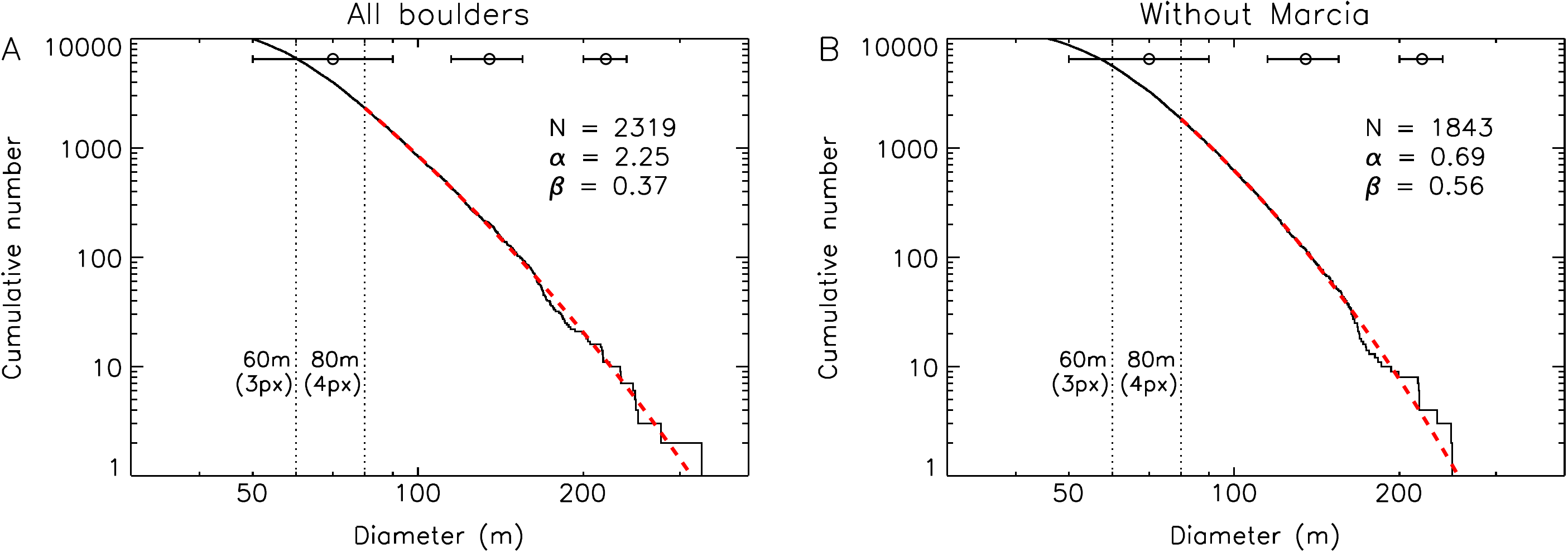}
\caption{Left-truncated Weibull distribution for Vesta boulders larger than 4~pixels, with ({\bf A}) and without ({\bf B}) Marcia boulders, displayed in cumulative format. The parameters of the best fit distribution (red curve) are listed. The error bars at the top indicate the uncertainty in boulder size at different diameters due to a 1~pixel measurement error.}
\label{fig:Weibull}
\end{figure}

\begin{figure}
\centering
\includegraphics[width=\textwidth,angle=0]{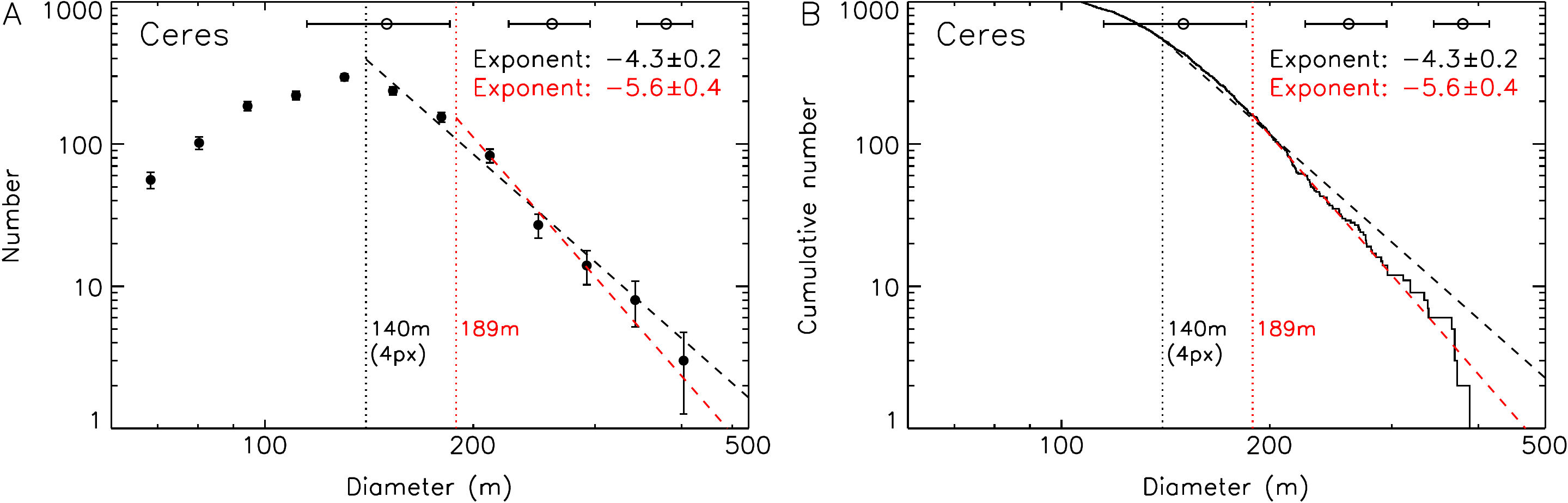}
\caption{Power law exponent for Ceres boulders. {\bf A}.~The differential distribution of all boulders counted on Ceres by \citeA{S18}. The black dashed line is the best-fit power law found when adopting a minimum boulder size of 4~pixels (black vertical dotted line). The red dashed line is the best-fit power law found with the minimum boulder size estimated by the ML method (red vertical dotted line). {\bf B}.~As (A) for the cumulative distribution. The error bars at the top indicate the uncertainty in boulder size at different diameters due to a 1~pixel measurement error.}
\label{fig:Ceres}
\end{figure}

\begin{figure}
\centering
\includegraphics[width=\textwidth,angle=0]{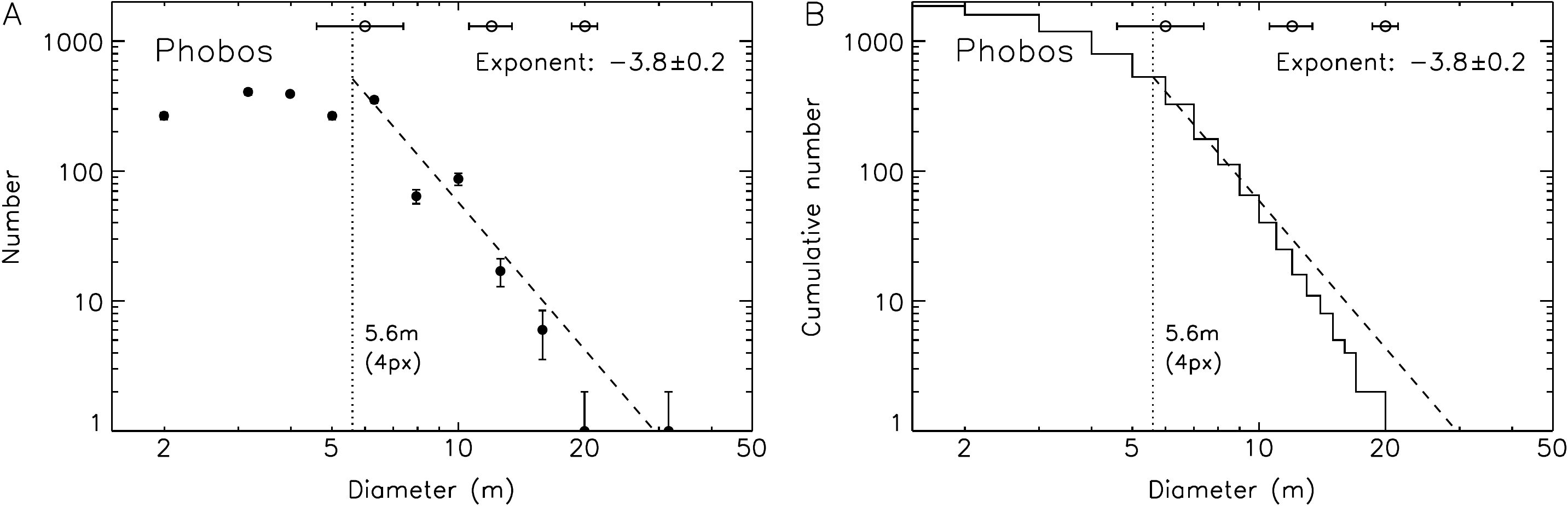}
\caption{Power law exponent for Phobos boulders. {\bf A}.~The differential distribution of all boulders counted on Phobos by \citeA{T00} in image 55103 (resolution 1.4~m per pixel). The black dashed line is the best-fit power law found when adopting a minimum boulder size of 4~pixels (black vertical dotted line). {\bf B}.~As (A) for the cumulative distribution. The error bars at the top indicate the uncertainty in boulder size at different diameters due to a 1~pixel measurement error.}
\label{fig:Phobos}
\end{figure}

\begin{figure}
\centering
\includegraphics[width=\textwidth,angle=0]{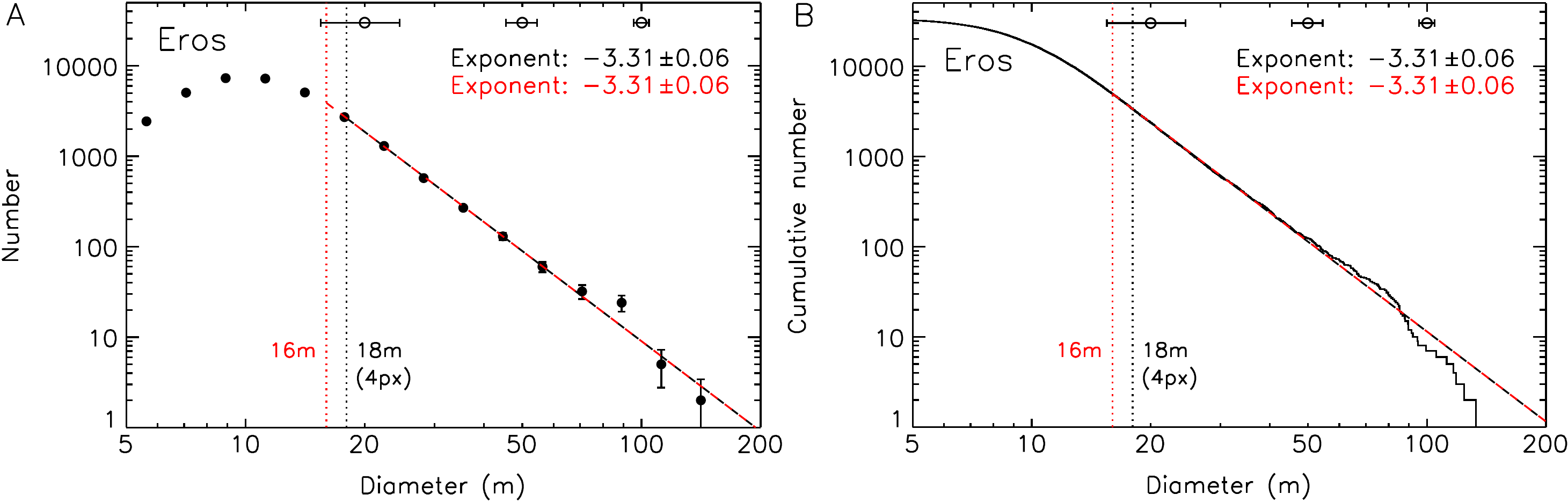}
\caption{Power law exponent for Eros boulders. {\bf A}.~The differential distribution of all boulders counted on Eros by \citeA{T01}. The black dashed line is the best-fit power law found when adopting a minimum boulder size of 4~pixels (black vertical dotted line), assuming a resolution of 4.5~m per pixel. The red dashed line is the best-fit power law found with the minimum boulder size estimated by the ML method (red vertical dotted line). {\bf B}.~As (A) for the cumulative distribution. The error bars at the top indicate the uncertainty in boulder size at different diameters due to a 1~pixel measurement error.}
\label{fig:Eros}
\end{figure}

\begin{figure}
\centering
\includegraphics[width=8cm,angle=0]{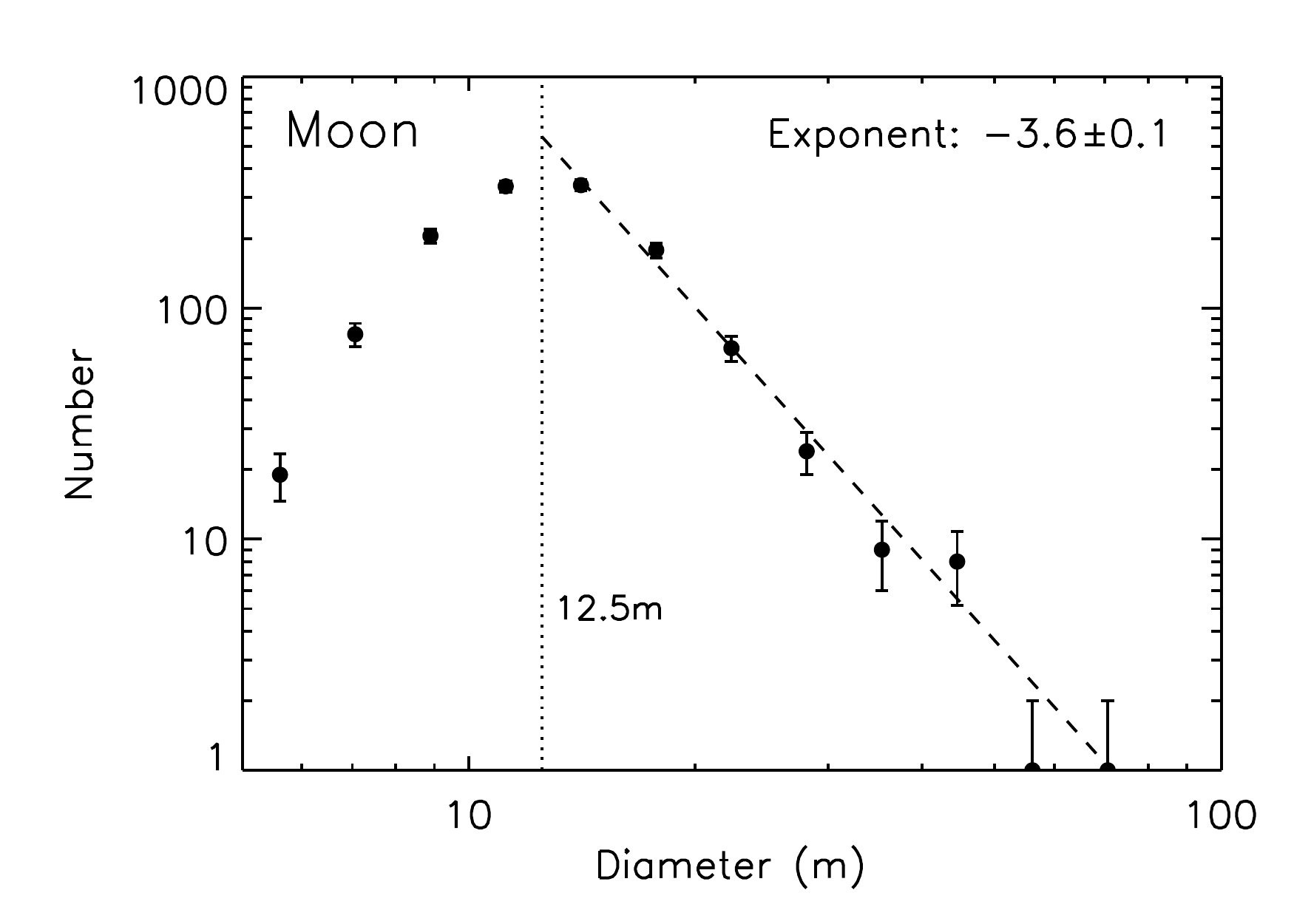}
\caption{Power law exponent for lunar boulders: Re-analyzing data for the Surveyor~VII site on the Moon from \citeA{C95}. The vertical dotted line is the minimum size we adopted, which is close to the quoted line pair resolution of 11~m. The dashed line is the best-fit power law.}
\label{fig:Moon}
\end{figure}

\begin{figure}
\centering
\includegraphics[width=8cm,angle=0]{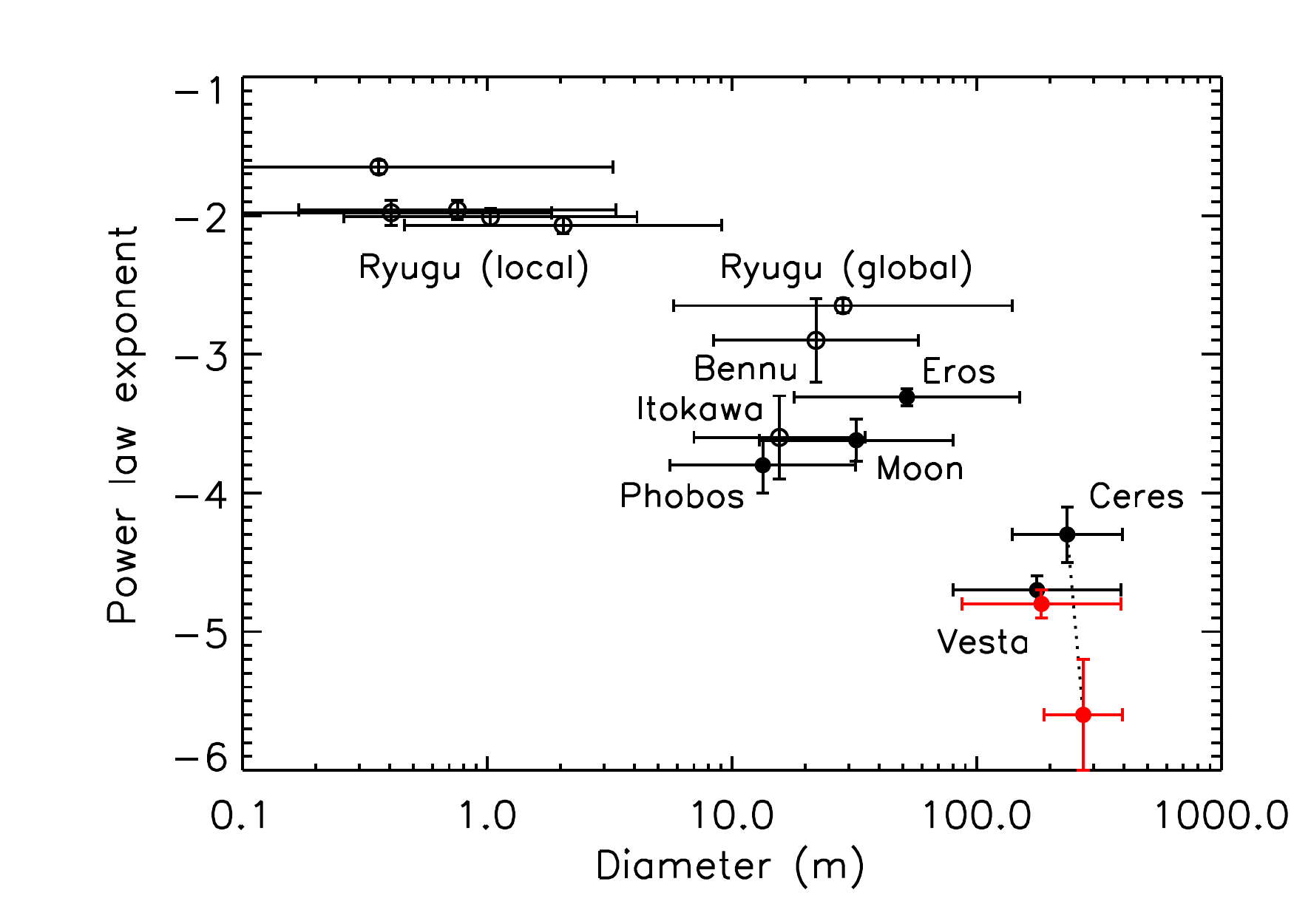}
\caption{Power law exponent for boulders on small Solar System bodies that were derived from populations of at least 100~boulders or particles (Table~\ref{tab:exponents}). The horizontal error bars indicate the size range over which the estimate was obtained. Open symbols are associated with suspected rubble pile asteroids. The black and red symbols for Vesta and Ceres refer to two ways of estimating $d_{\rm min}$ for the ML fit.}
\label{fig:power_law_exponents}
\end{figure}

\end{document}